\documentclass[aps, pre, amsmath, amssymb, showpacs,showkeys,floatfix,superscriptaddress,twocolumn]{revtex4}

\usepackage{natbib}
\usepackage{graphicx}
\usepackage{bm}
\usepackage{color}
\usepackage{epic, eepic}
\usepackage{subfigure}

\newcommand{\ddt}[1]    {\partial_t {#1} }
\newcommand{\dddt}[1]   {\frac{d {#1}}{d t}}   
\newcommand{\dddthat}[1]{\frac{d {#1}}{d \hat{t}}}   
\newcommand{\ddx}[2]    {\partial_{#1} {#2} }
\newcommand{\ddxi}[1]   {\partial_i {#1} }
\newcommand{\ddxj}[1]   {\partial_j {#1} }
\newcommand{\ddxsq}[2]  {\partial_{#1}^2 {#2} }
\newcommand{\ddxjsq}[1] {\partial_j^2 {#1} }

\newcommand{\spav}[1]    {\langle {#1} \rangle}
\newcommand{\volav}[1]  {\langle {#1} \rangle_V}
\newcommand{\plav}[1]   {\langle {#1} \rangle_A}
\newcommand{\hav}[1]    {\langle {#1} \rangle_H}

\newcommand{\yav}[1]    {\langle {#1} \rangle_y}
\newcommand{\tav}[1]    {\langle {#1} \rangle_t}
\newcommand{\ensav}[1]  {\langle {#1} \rangle}

\newcommand{\abs}[1] {\left| {#1} \right|}

\newcommand{\av}[1]     {\overline{#1}}
\newcommand{\fl}[1]     {#1'}
\newcommand{\sav}[1]      {\widetilde{#1}}
\newcommand{\sfl}[1]      {\fl{#1}}

\newcommand{\bA}[0]     {\mathcal{A}}
\newcommand{\bD}[0]     {\mathcal{D}}
\newcommand{\bP}[0]     {\mathcal{P}}
\newcommand{\bB}[0]     {\mathcal{B}}
\newcommand{\bR}[0]     {\mathcal{R}}

\renewcommand{\Re}[0]     {\mathrm{Re}}
\newcommand{\Ra}[0]     {\mathrm{Ra}}
\newcommand{\Nu}[0]     {\mathrm{Nu}}
\renewcommand{\Pr}[0]     {\mathrm{Pr}}

\renewcommand{\vec}[1]  {{ \bf #1}}

\newcommand{\symop}     {S}
\newcommand{\trans}     {\vec{d}}

\newcommand{\ignore}[1]{}

\begin{document}

\title{Wind and boundary layers in Rayleigh-B\'{e}nard convection. \\ 
I. Analysis and modeling}

\author{Maarten \surname{van Reeuwijk}}

\email{m.vanreeuwijk@imperial.ac.uk}

\affiliation{Department of Civil and Environmental Engineering, Imperial College London, Imperial College Road, London, SW7 2AZ, United Kingdom}

\author{Harm J.J. \surname{Jonker}}

\affiliation{Department of Multi-Scale Physics and
             J.M. Burgers Center for Fluid Dynamics,
             Delft University of Technology,
             Lorentzweg 1, 2628 CJ Delft, The Netherlands}

\author{Kemo \surname{Hanjali\'{c}}}

\affiliation{Department of Multi-Scale Physics and
             J.M. Burgers Center for Fluid Dynamics,
             Delft University of Technology,
             Lorentzweg 1, 2628 CJ Delft, The Netherlands}

\affiliation{Department of Mechanics and Aeronautics, University of Rome, ``La Sapienza'', Rome, Italy}

\date{December 24, 2007}



\begin{abstract}
The aim of this paper is to contribute to the understanding and to model the processes controlling the amplitude of the wind of Rayleigh-B\'{e}nard convection. 
We analyze results from direct simulation of an $L/H=4$ aspect-ratio domain with periodic sidewalls at $\Ra=\{10^5, 10^6, 10^7, 10^8 \}$ and at $\Pr=1$ by
decomposing independent realizations into wind and fluctuations.
It is shown that deep inside the thermal boundary layer, horizontal heat-fluxes exceed the average vertical heat-flux by a factor 3 due to the interaction between the wind and the mean temperature field.
These large horizontal heat-fluxes are responsible for spatial temperature differences that drive the wind by creating pressure gradients. 
The wall fluxes and turbulent mixing in the bulk provide damping.
Using the DNS results to parameterise the unclosed terms, a simple model capturing the essential processes governing the wind structure is derived.
The model consists of two coupled differential equations for wind velocity and temperature amplitude.
The equations indicate that the formation of a wind structure is inevitable due to the positive feedback resulting from the interaction between the wind and temperature field.
Furthermore, the wind velocity is largely determined by the turbulence in the bulk rather than by the wall-shear stress.
The model reproduces the $\Ra$ dependence of wind Reynolds number and temperature amplitude.
\end{abstract}

\maketitle

\section{Introduction}

One of the characteristic features of Rayleigh-B\'{e}nard convection is a large scale circulation or 'wind, which is generated autonomously by the system and is of great importance for the effectivity of the heat transfer \citep{Grossmann2000}. Although first observed in a large aspect-ratio $\Gamma = L/H$ cell \citep{Krishnamurti1981}, the wind has been studied mostly in smaller aspect-ratio cells \citep{Lam2002,Niemela2001,Qiu1998,Qiu2000,Sreenivasan2002,Wang2003,Xi2004,Xin1997,Xin1996, Kadanoff2001}.
The wind has complex dynamics, in that it changes its direction erratically at timescales far exceeding the convective turnover time \citep{Niemela2001,Sreenivasan2002}.
In the case of cylindrical cells, there are two separate ways for reversals to occur \cite{Brown2005, Brown2006}.
First, the wind structure can change its orientation by rotating in the azimuthal direction, which leads to reversals if the system rotates over $180^o$. 
The second mechanism for reorientation is by cessation, when the large scale circulation briefly halts and restarts with a different random orientation.
The wind dynamics change depending on the aspect-ratio $\Gamma$ and the Rayleigh number $\Ra$.
In cylindrical $\Gamma=1/2$ domains, the wind structure (normally one roll throughout the entire domain) breaks up into two counter-rotating rolls on top of each other \citep{Verzicco2003} around $\Ra=10^{10}$. 
At even higher $\Ra$, roughly around $10^{12}$, the wind substantially weakens
\citep{Niemela2003, Amati2005, Niemela2006}.
For large aspect-ratio domains, the wind structure tends to be weaker relative to the fluctuations \citep{Kerr1996, Hartlep2005, Niemela2006, Shishkina2006, Verdoold2006}.

\ignore{
Experiments indicate that the scaling exponents such as the Nusselt number $\Nu$ and the Reynolds number $\Re$ do not depend on $\Gamma$, but that fluctuations are much stronger relative to the wind than in small aspect-ratio cells.
}

Several models have been developed recently to explain the complex long-term dynamics of the wind, in particular the wind reversals and reorientations.
The first model to explain wind-reversals was by Sreenivasan  \emph{et al.}\ \citep{Sreenivasan2002}, which is based on the conceptual image of a double-well potential representing the preference for an average clock-wise or counter-clockwise motion.
The turbulence is modelled by stochastic fluctuations, which are responsible for sudden reversals when strong enough to overcome the energy barrier separating the two states.
A different approach was taken by Fontenele Araujo \emph{et al.} \citep{Araujo2005}, who derived a deterministic model describing the dynamics of a thermal on a circular trajectory in a linearly unstably stratified fluid.
The resulting equations are similar to the Lorenz equations and exhibit chaotic flow reversals in a specific region of the $\Ra$-$\Pr$ phase space.
The two-dimensional models described above can only reproduce reversals by cessations, and do not facilitate reorientation by rotations, which occur more often \citep{Brown2005, Brown2006}. 
Brown and Ahlers \citep{Brown2007} recently presented a model which is capable of predicting reorientations both by rotations and cessations.
This model is inspired by the Navier-Stokes equations and constitutes two stochastical differential equations, one for the temperature amplitude and one for the azimuthal orientation.

Despite these significant advances in the understanding of the long-term wind-dynamics, it is currently not clear exactly how the wind is driven and how the turbulence and wall-fluxes influence the wind amplitude.
It is known that the wind is sustained by the spatial differences in mean temperature along the sidewalls \citep{Burr2003}.
However, it is not clear what generates these temperature differences, and what the relation between the temperature differences and the wind velocity is.
\ignore{
move to discussion?

Consequently, models need to make assumptions.
Fontenele Araujo \emph{et al.} \citep{Araujo2005} assume that buoyancy and drag dominate the plume behavior.
The drag term is parameterised with a relation which is valid both in the laminar and turbulent regime.
Brown and Ahlers \cite{Brown2007} assume that the temperature amplitude is proportional to the wind velocity amplitude, as both are measures for the strength of the wind.
}
In this paper, we use direct numerical simulation of a rectangular $\Gamma=4$ domain at $\Pr=1$ and $\Ra=\{10^5, 10^6, 10^7, 10^8 \}$ with periodic lateral boundary conditions to provide insight into these questions.
We derive a model for the wind based on the Reynolds-averaged Navier-Stokes equations, which consists of two coupled ordinary differential equations for the average wind velocity and temperature amplitude.
This simple conceptual model provides insight in the role of turbulence in the bulk on the wind velocity and the neccessity for a wind-structure to develop.
In the accompanying paper \cite{vanReeuwijk2007c}, we will focus on the boundary layers at the top and bottom walls and their interaction with the wind, and propose new scaling relations for $\lambda_u$ and $C_f$.

The paper is organized as follows.
The governing equations, averaging strategies and their relation to the system's symmetries are discussed in section \ref{par:theory}.
The method of wind extraction by symmetry-accounting ensemble-averaging is outlined in section \ref{par:symav}.
Similar to domains with sidewalls, a wind structure develops for unconfined domains \cite{Kerr1996, Hartlep2003, Roode2004, Hartlep2005, vanReeuwijk2005}.
As the wind structure is not kept in place by side walls, it can be located anywhere in the domain because which complicates extracting the wind structure.
However, by identifying the wind structure and proper alignment of different realizations (by accounting for symmetries), a wind structure can also be unambiguously defined for unbounded domains \cite{vanReeuwijk2005}.
Details about the code and simulations are discussed in section \ref{par:simdetails}.
Some results of $\Nu$ and $\Re$ as a function of $\Ra$ are presented in section \ref{par:classical}.
The wind and the temperature field following from the symmetry-accounted
averaging are presented in section \ref{par:symavresults}.
The decomposed profiles of kinetic energy are presented in section \ref{par:keprofiles}, eliciting the importance of the wind for the dynamics of the flow.
It turns out that the wind structure has a significant influence on the
redistribution of heat in the system, as is discussed in section
\ref{par:heatredistribution}.
Section \ref{par:windmechanism} contains a discussion how the wind is maintained by a study of the momentum and temperature budgets at several positions of the flow, and a detailed feedback mechanism is sketched.
Then, the findings are synthesized in a simple conceptual model in section \ref{par:windmodel}, and conclusions are presented in section \ref{par:conclusions}.

\section{Background}

\subsection{\label{par:theory}Theory}

Rayleigh-B\'{e}nard convection is generated when a layer of fluid with
thickness $H$ between two parallel plates is subjected to a positive
temperature difference $\Delta \Theta$ between top and bottom plate.
The positive temperature difference causes the buoyant fluid to become unstable,
causing convection and thereby enhancing the heat-transport through the layer.
In the dynamics one can observe organized motion such as plumes, jets and wind \citep{Kadanoff2001}.
For an incompressible Boussinesq fluid with isobaric thermal expansion
coefficient $\beta$, viscosity $\nu$ and thermal diffusivity $\kappa$, the
governing equations are
\begin{gather}
\label{eq:NS}
\ddt{u_i} + \ddxj{u_j u_i} = - \rho^{-1} \ddxi{p} + \nu \ddxjsq{u_i} +
                             \beta g \Theta \delta_{i3}, \\
\label{eq:T}
\ddt{\Theta} + \ddxj{u_j \Theta} = \kappa \ddxjsq{\Theta}, \\
\label{eq:div}
\ddxj{u_j} = 0.
\end{gather}
Here $\rho$ is the density, $g$ the gravitational constant, $u_i$ represents
the fluid velocity, $\Theta$ the temperature and $p$ the pressure.
No-slip velocity and fixed temperature are enforced on the top- and bottom
walls.
The problem can be characterized by the Prandtl number $\Pr = \nu
\kappa^{-1}$ which represents the ratio of viscosity and thermal diffusivity
and the Rayleigh number $\Ra= \beta g \Delta \Theta H^3 (\nu \kappa)^{-1}$ 
which relates the buoyant and viscous forces.
The system reacts by convective motion characterized by the 
Reynolds number $\Re = U H \nu^{-1}$ and by an enhanced heat
transfer through the Nusselt number $\Nu = \phi H (\kappa \Delta
\Theta)^{-1}$ which is the non-dimensional heat-flux through the top and bottom
wall.
Here $U$ is a characteristic velocity and $\phi$ the heat-flux.
Both $\Re$ and $\Nu$ are unknown \emph{a priori}.

Since definitions for the processes occurring in Rayleigh-B\'{e}nard
convection are not unambiguous, a small glossary is given here.
We prefer to use the term \emph{wind structure}, which generalizes the
terms wind and large scale circulation, in that it involves both the velocity
and the temperature field.
This wind structure normally features convection rolls, which are the quasi-steady roll-like structures.
Thermals and plumes are the unsteady structures erupting from the boundary layers and propagating into the bulk.
Spatial averages will be denoted by $\volav{}$, $\plav{}$ and
$\hav{}$ for volume-, plane- and height-averaging, respectively.
The plane-average is in the homogeneous (x and y) directions.
Time and ensemble averages will be denoted by $\tav{}$ and $\ensav{}$.

In what follows a domain of size $L \times L \times H$ with $L = \Gamma H$ and
$\Gamma$ the aspect-ratio will be considered.
Periodic boundary conditions are imposed on the side walls.
Applying $\plav{}$ to the incompressibility constraint (\ref{eq:div}) and
using impermeability at the top and bottom wall yields that the
plane-averaged velocities $\plav{u}=\plav{v}=\plav{w}=0$.
Taking the ensemble average of the temperature equation (\ref{eq:T}) and
the fixed temperature boundary conditions gives after some manipulation that
\begin{equation}
  \label{eq:Nu}
  \Nu = \frac{H}{\kappa \Delta \Theta}
               \left( \ensav{\fl{w} \fl{\Theta}} -
                      \kappa \ddx{z}{\ensav{\Theta}}
               \right),
\end{equation}
which states that the mean total heat-flux is constant in the vertical and
directly related to $\Nu$.

Interesting differences exist in the standard way of averaging between
experiments, simulation and theory.
We focus on laterally unbounded domains or domains with periodic boundary condition and will use the overbar $\av{X}$ to denote a generic averaging operator.
Experiments normally employ the time-average $\tav{X}$ and theory
the ensemble average $\ensav{X}$.
In simulations of unbounded Rayleigh-B\'{e}nard convection it is customary to use a plane-average $\plav{X}$, because it can be evaluated at every time instant.
The underlying assumption is that $\av{X}$ coincides with the ensemble average
$\ensav{X}$ and the time average $\tav{X}$, but there are some subtleties
that require attention here.
It can be imagined that $\plav{X}$ will approach $\ensav{X}$ for $\Gamma$
sufficiently large, as a typical realization is expected to be of size $O(H)$ by
which the domain would contain roughly $\Gamma^2$ of those realizations.
The time average $\tav{X}$ produces one independent realization every $O(t^*)$
with $t^*=H/U$ the typical timescale, and it can be expected that for averaging
over sufficiently long times it converges to the ensemble average so that
$\tav{X} = \plav{X} = \ensav{X}$.
However, this presumes that the system's phase space is not partitioned, i.e
that the system will visit all its possible states within finite time.
When this condition is satisfied the system is ergodic, and this is one of the
primary assumptions underlying turbulence theory \citep{Frisch1995,
Galanti2004}
From the continuity equation, it follows that $\plav{u_i} = 0$, by which all natural averages, i.e.
long-time, ensemble and spatial averages vanish as $\av{u}=\av{v}=\av{w}=0$. Hence one would conclude that Rayleigh-B\'{e}nard convection is comprised purely of fluctuations, which is in conflict with the ubiquitous large scale circulation or wind.

The paradox of the existence of a mean wind and the restriction
of $\av{u}=\av{v}=\av{w}=0$ can be resolved by taking into account the
symmetries of the problem \citep{vanReeuwijk2005}.
When there are symmetries in the domain, there is a chance for symmetric conjugate modes (such as clockwise and counter-clockwise mean flow in the cell) to cancel each other, given enough time (through wind-reversals) or realizations.
By accounting for symmetries before performing ensemble-averaging, all fields are properly 'aligned' before the averaging takes place, allowing the modes that would normally be cancelled by their symmetric conjugates to persist.
The resulting average field of velocity and temperature is the wind
structure and in the fluctuations are the actions of the plumes.

\ignore{
For unbounded (periodic) domains it is usual to average results over the homogeneous directions, so that the ensemble average $\ensav{u_i}=0$, and the velocity-field seems comprised of fluctuations only.
A decomposition into a wind structure and fluctuations gives a physically more accurate representation.
Indeed, it will be shown the wind-structure is part of a significant portion of the horizontal variance, and that the characteristic peak in the profile of kinetic energy by which the boundary layer thickness is normally identified, are due to the wind structure.
}
\subsection{\label{par:symav}Symmetry-accounted ensemble-averaging}

\begin{figure*}
\centering
\subfigure[Classical averaging]{\includegraphics[width=55mm]{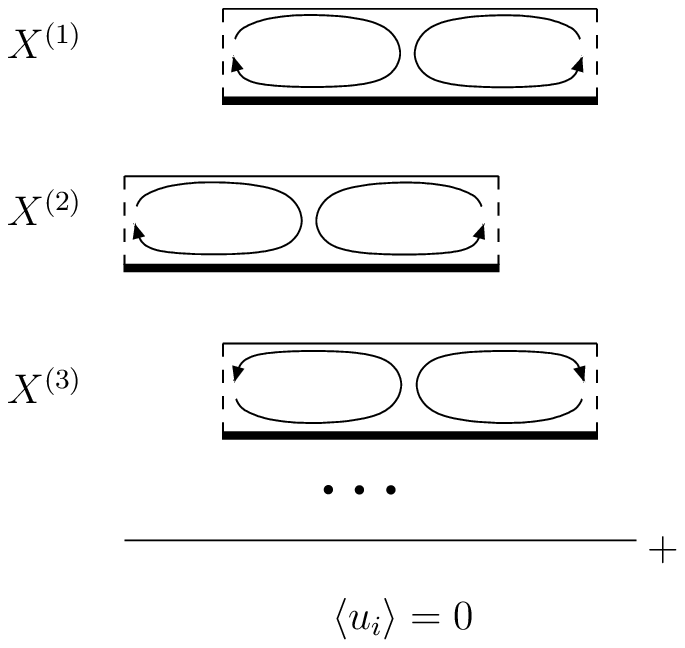}}
\hspace{15mm}
\subfigure[Symmetry-accounted  averaging]{\includegraphics[width=64mm]{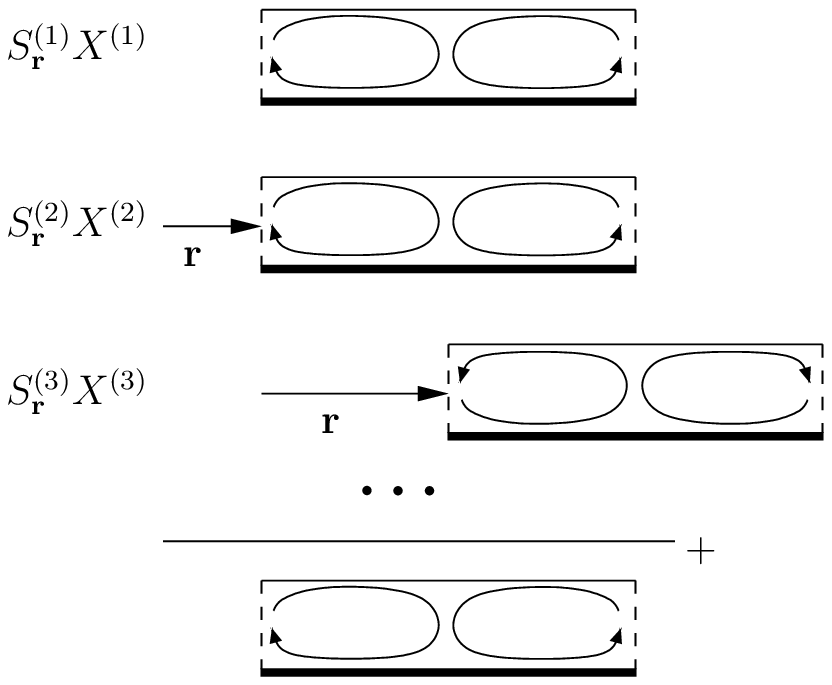}}
\caption{\label{fig:symav} Ensemble averaging in domains with periodic
side walls. a) Classical averaging results in zero mean wind; b) When accounting
for symmetries by translating the realizations if necessary, the wind structure
is preserved.}
\end{figure*}

The rationale of symmetry-accounted ensemble-averaging has been presented for
general domains elsewhere \citep{vanReeuwijk2005} and we discuss here only the
application to our case with periodic side walls.
The system has two symmetries: a discrete rotational symmetry and a 
continuous translational invariance.
The most important symmetry to take into account here is the translational 
invariance in $x,y$.
When considering an ensemble of realizations $\{ X^{(1)},X^{(2)}, \ldots, X^{(N)} \}$, it can be expected that a wind structure is present in all of them, although its location will differ per realisation.
When one takes the average of this ensemble, the wind structure will be averaged
out so that nothing but fluctuations remain (Fig.\ \ref{fig:symav}a).
However, due to the translational invariance, one can translate a 
realization and obtain another valid solution to the equations.
By translating each realization $X^{(\alpha)}$ over a distance $\trans^{(\alpha)}$ such that the wind structures become aligned, the averaging out of the wind can be prevented, as is sketched in Fig.\ \ref{fig:symav}b.

The translational operator can be denoted by $\symop_{\trans}$ with 
$\trans \equiv (d_x, d_y)$ representing the relative displacement. 
Operating on a field $X$, the translational operation is simply $\symop_{\trans} X = X(x-d_x, y-d_y, z)$.
Symmetry-accounted averaging then, means to translate each realization $\alpha$ before averaging as 
\begin{equation}
  \label{eq:symav}
  \begin{split}
  \sav{X} &= \sum_{\alpha=1}^{N} \symop_{\trans}^{(\alpha)} X^{(\alpha)} \\
          &= \sum_{\alpha=1}^{N} X^{(\alpha)}(x-d_x^{(\alpha)}, y-d_y^{(\alpha)}, z),
  \end{split}
\end{equation}
where $\trans^{(\alpha)}$ is chosen such that the wind structure does not average out.
An alternative way to look at symmetry-accounted ensemble-averaging is that it involves a preprocessing step before performing the ensemble-averaging.
The fluctuating field is defined as
\begin{equation}
  \sfl{X}^{(\alpha)} = X^{(\alpha)}(x-d_x^{(\alpha)}, y-d_y^{(\alpha)}, z) - \sav{X}(x,y,z),
\end{equation}
and it is straightforward to prove that $\sav{\sfl{X}} \equiv 0$.
Hence, the results can be interpreted exactly the same way as those from classical Reynolds-decomposition.

The symmetry-accounted ensemble average $\sav{X}$ is closely related to the classical (ensemble, long-time or spatial) average $\av{X}$, and we will point out some useful relations between the two.
Due to translation invariance all statistics $\av{X}$ are a function of $z$ only, whereas the symmetry-accounted average $\sav{X}$ retains the full three-dimensional structure.
The first important relation is that the plane-average of the symmetry-accounted
average is identical to the classical average as
\begin{equation}
\label{eq:symav_mean}
\plav{\sav{X}} = \av{X}
\end{equation}
which follows directly from substitution of the two different decompositions $X = \sav{X}(x,y,z) + \sfl{X}(x,y,z)$ and $X = \av{X}(z) + X"(x,y,z)$ into the expression $\plav{X}$.
The second useful relation pertains the variance, and is given by
\begin{equation}
\label{eq:symav_variance}
\plav{\sav{X} \sav{X}} + \plav{\sav{\fl{X}\fl{X}}} = 
\av{X}~\av{X} + \av{X" X"} 
\end{equation}
which can be obtained similarly.
Expression \eqref{eq:symav_variance} is particularly useful for the analysis of the profiles of kinetic energy (\ref{par:keprofiles}) and for the decomposed vertical heat-fluxes (section \ref{par:heatredistribution}).

If the wind structure was known \emph{a priori}, the displacement $\trans$ would be the only unknown per realization, and (\ref{eq:symav}) could be applied immediately.
Unfortunately this is not the case, as both the wind structure and $\trans$ 
are unknown.
Therefore, an iterative technique is used by which the wind structure and the displacements are determined simultaneously, gradually improving the estimation for the wind structure in successive iterations \citep{Eiff1997}.
The only assumption needed for this method is that -- among the majority of
the realizations -- only one persistent structure (mode) is present 
inside the domain.

To start the iterative process a reference field $X_0(\vec{x})$ is needed, for which an arbitrarily picked realization is used -- the wind structure is present in every realization so the starting point should not make a difference.
Using a cross-correlation function $C(X, Y)$, every realization can be compared to $X_0(\vec{x})$, and the location of maximum correlation is picked as the
displacement vector:
\begin{equation}
\label{eq:displacement}
\vec{\trans}^{(\alpha)} \leftarrow \max_{\vec{r}}\ C (\symop_{\vec{r}} X^{(\alpha)}, X_0).
\end{equation}

There is some freedom in choosing how to calculate the overall 2D (in $x$ and $y$) correlation field, as it can be constructed from any combination of the three-dimensional fields $X \in \{u_i,\Theta, p\}$.
In this case we opted for the instantaneous height-averaged temperature $\hav{\Theta}$ which is closely related to the wind structure as $\hav{\Theta} > 0$ where $w>0$ and vice versa.
Denoting the reference field by $X_0(x,y)=\hav{\Theta_0}$ and a different realisation by $Y(x,y)$, the cross-correlation function is given by
\begin{equation}
  C(\symop_{\vec{r}} Y, X_0) = \frac{ \iint Y'(x-r_x, y-r_y) X_0'(x, y) dx dy }
                 {\sigma_{X} \sigma_{Y} }.
\end{equation}
Here, $X_0' = X_0 - \plav{X_0}$ and $Y' = Y - \plav{Y}$ are the deviations from the mean, and $\sigma_X$ and $\sigma_Y$ are the standard deviations of $X_0$ and $Y$.
The displacement vector $\vec{d}$ is just the coordinate pair $(r_x, r_y)$ for which the correlation is maximal.
For computational efficiency, the correlation is determined via FFT's.
After calculating $\vec{d}^{(\alpha)}$ for all realizations, a new and improved
estimation can be determined by
\begin{equation}
\label{eq:iter}
\sav{X}_{n+1} = \frac{1}{N} \sum_{\alpha=1}^{N} X^{(\alpha)}(x-d_x^{(\alpha)}, y-d_y^{(\alpha)}, z)
\end{equation}
Repeatedly applying \eqref{eq:displacement} and \eqref{eq:iter} with $X_0$ replaced with $X_n$ and until $\sav{X}_{n+1} = \sav{X}_{n}=\sav{X}$ results in the wind structure, or symmetry accounted average, as well as the relative displacements $\vec{d}^{(\alpha)}$.
It is emphasized that vertically averaged fields are only used to determine the relative displacements $\vec{d}^{(\alpha)}$; the resulting wind structure is fully three-dimensional.
\subsection{\label{par:simdetails}Simulation details}

Direct numerical simulation (DNS) is used to generate the independent
realizations for the symmetry-accounted averaging.
The code is based on finite volumes and has the equations (\ref{eq:NS}-\ref{eq:div})
discretized and implemented on a staggered grid.
Central differences are used for the spatial derivatives and time integration
is by a second order Adams-Bashforth scheme.
The code is fully parallelized and supports grid clustering in the wall-normal
direction. 
Special attention has been given to conservation of variance by preserving the
symmetry-properties of the discrete advective and diffusive operators 
\citep{Verstappen2003}.
Further details of the code can be found elsewhere \citep{vanReeuwijk2007}.

\begin{figure}
  \center
  \setlength{\unitlength}{1mm}
  \includegraphics*[width=75mm]{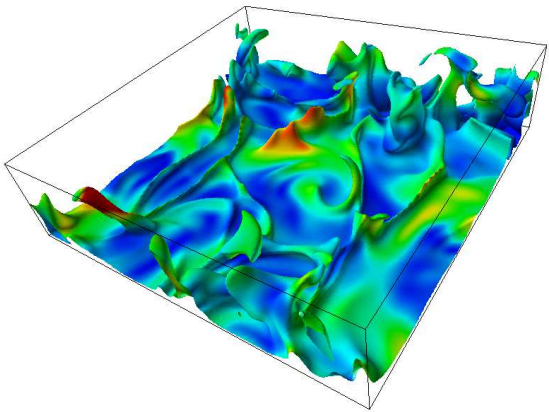}
  \caption{\label{fig:inst3d} (Color online) Snapshot from one of the direct numerical
simulations at $\Ra=10^6$ and $\Pr=1$. Shown is an iso-surface of temperature,
colored by the kinetic energy.}
\end{figure}

Resolving all the length-scales makes direct numerical simulation a powerful
research tool, as one has the complete four-dimensional solution of the
Navier-Stokes equations at hand.
However, DNS is limited to relatively low $\Re$ as the computational demands
quickly become prohibitive, scaling approximately as $\Re^{3}$.
Furthermore, both the thermal and hydrodynamic boundary layer, $\lambda_\Theta$
and $\lambda_u$ respectively, should be fully resolved as undersampling will
lead to overestimation of $\Nu$ \citep{Kerr1996}.

Simulations have been performed at $\Pr=1$ and
$\Ra = \{ 10^5, 10^6, 10^7, 10^8 \}$ for an aspect-ratio $\Gamma=L/H=4$ domain.
The grid resolution and other relevant information is given in Table
\ref{tab:simdetails}.
The Reynolds number $\Re$ has been obtained from the peak of $\av{\fl{u} \fl{u}}$ and $\Re_\tau = u_\tau H / \nu$, with $u_\tau = \sqrt{ \nu \frac{d}{dz} k^{1/2}}$ at the wall.
Here, $k$ represents the turbulent kinetic energy, which may not be the most
ideal approximation of the shear velocity; normally the mean horizontal
velocity is used.
However, from the 'classical' (ensemble-average) point of view, there is no mean wind so that the only available data is from fluctuations.

The grid clustering in the near-wall region has been chosen such that on average 8 cells were present in the thermal boundary layer.
The kinetic boundary layer which is thicker  than the thermal boundary layer at $\Pr=1$, contained about 16 cells on average. 
A snapshot of one of the simulations at $\Ra=10^6$ clearly shows the unstable sheet-like plumes emerging from the boundary layers (Fig. \ref{fig:inst3d}).
Ten independent simulations with slightly perturbed initial conditions have been performed for all but the highest $\Ra$, as the
computational demands were too high.
At $\Ra=10^8$ on the $640^2 \times 320$ grid, one convective turnover time took
2500 hour on one SGI Origin 3800 processor and even with 128 processors this is
20 wall-clock hours per turn-over time.

\renewcommand\arraystretch{2}
\renewcommand\tabcolsep{2mm}
\begin{table*}
\centering
\caption{
\label{tab:simdetails}
Simulation details}
\begin{tabular}{cccccccc}

$\Ra$ & grid & $\Delta t / t^* \times 10^3$ & $T / t^*$ & \#sims & 
$\Nu$ & $\Re$ & $\Re_\tau$ \\
$1.15 \times 10^5$ & $128^2 \times 64$  & 1.13 & 68 & 10 & 4.5 & 54 & 32 \\ 
$1.0 \times 10^6$ & $192^2 \times 128$ & 0.57 & 20 & 10 & 8.3 & 157 & 70 \\ 
$1.0 \times 10^7$ & $256^2 \times 256$ & 0.45 & 20 & 10 & 16.1 & 458 & 160 \\
$1.0 \times 10^8$ & $640^2 \times 320$ & 0.11 & 5 & 1 & 31.1 & 1499 & 210
\end{tabular}
\end{table*}
\renewcommand\arraystretch{1}
\section{\label{par:classical}Classical results}

\begin{figure}
   \subfigure[]{\includegraphics[width=90mm]{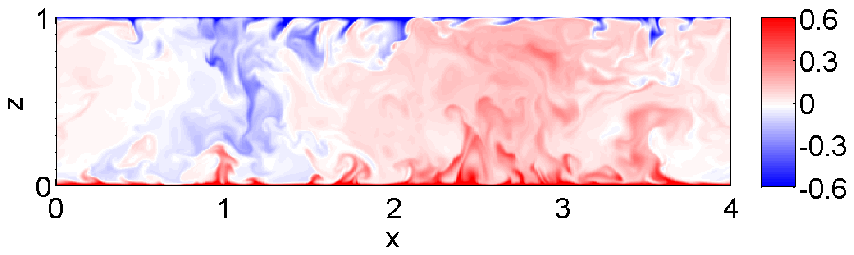}}
   \subfigure[]{\includegraphics[width=90mm]{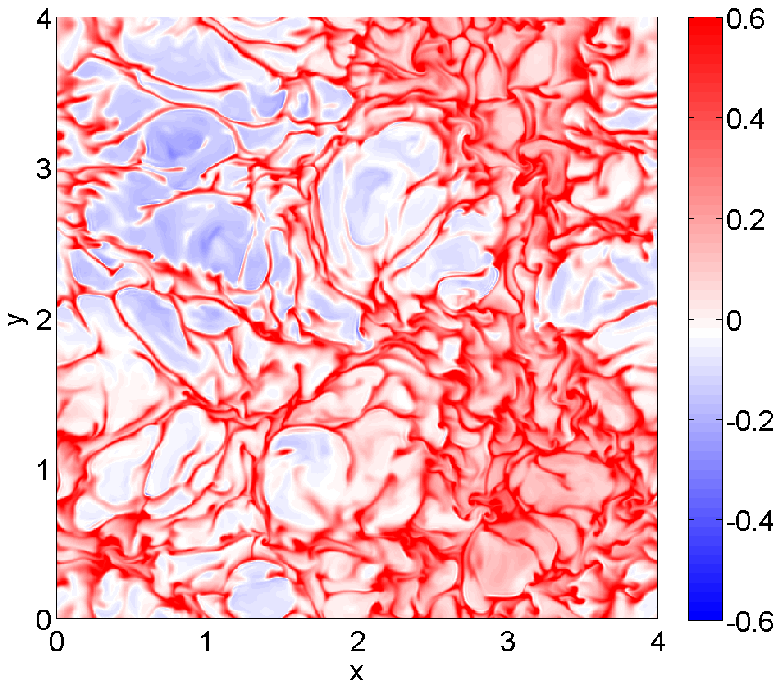}}
\caption{\label{fig:movies}(Color and movies online) Cross-sections of the temperature field at $\Ra=10^8$ and $\Pr=1$. a) An $x-z$ cross-section. b) An $x-y$ cross-section at the edge of the bottom thermal boundary layer. The online movies are accelerated 5 times, and blue and red represent low and high temperatures, respectively.}
\end{figure}

Instantaneous cross-sections of the temperature field are shown in Fig. \ref{fig:movies} at $\Ra=10^8$. The dynamic behavior can be viewed in the online animations
\footnote{See EPAPS Document No. [number will be inserted by publisher] for the dynamical behavior of the temperature field at $\Ra=10^8$ and $\Pr=1$ in the boundary layer and in a vertical cross-section. For more information on EPAPS, see http://www.aip.org/pubservs/epaps.html.}
The vertical (x-z) cross-section of the temperature field (Fig.\ \ref{fig:movies}a) clearly shows the spatial segregation of hot areas where upward thermals dominate and cool areas where the downward thermals dominate.
Fig.\ \ref{fig:movies} shows a horizontal (x-y) cross-section of the temperature field at the edge of the thermal boundary layer.
The boundary layer is a network of sheet-like plumes, which is coarse where the average flow is downward and dense where it is upwards.
The sheets are formed by impingement of cold plumes onto the plate, as the hot fluid in the boundary layer is pushed away.
These hot sheets move towards the region with ascending flow, where they seem to form an ever-contracting network of plumes.
Where the network is dense, the plumes detach and the average flow is upward.

\ignore{
\begin{figure*}
  \center
  \subfigure[]{\includegraphics*[width=60mm]{fig_t01_bl_Ra1e6_Pr1.eps}}
  \hspace{1cm}
  \subfigure[]{\includegraphics*[width=60mm]{fig_t01_center_Ra1e6_Pr1.eps}}
  \subfigure[]{\includegraphics*[width=60mm]{fig_t01_bl_Ra1e7_Pr1.eps}}
  \hspace{1cm}
  \subfigure[]{\includegraphics*[width=60mm]{fig_t01_center_Ra1e7_Pr1.eps}}
  \subfigure[]{\includegraphics*[width=60mm]{fig_t01_bl_Ra1e8_Pr1.eps}}
  \hspace{1cm}
  \subfigure[]{\includegraphics*[width=60mm]{fig_t01_center_Ra1e8_Pr1.eps}}
  \caption{\label{fig:insttemp} Cross-sections of temperature in the
thermal boundary layer (a,c,e) and center (b,d,f) for various $\Ra$. Figs.\
(a,b): $\Ra=10^6$; (c,d): $\Ra=10^7$; (e,f):$\Ra=10^8$.}
\end{figure*}
}

\begin{figure*}
  \centering
  \subfigure[]{\includegraphics{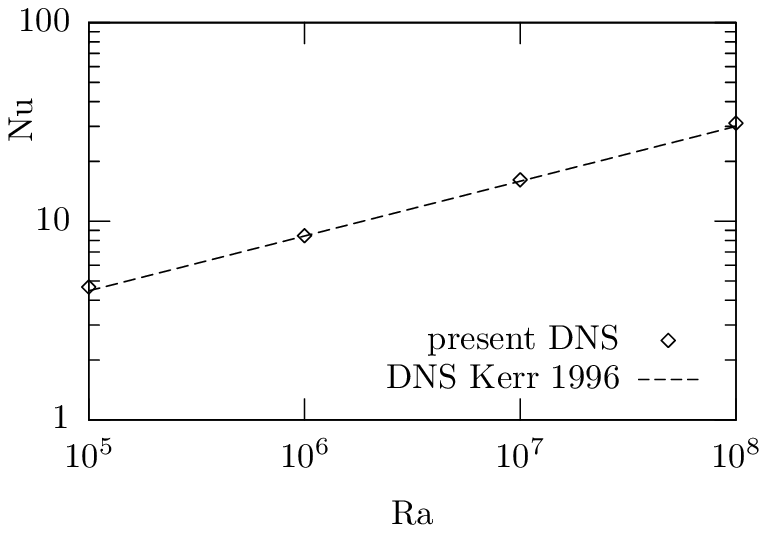}}
  \subfigure[]{\includegraphics{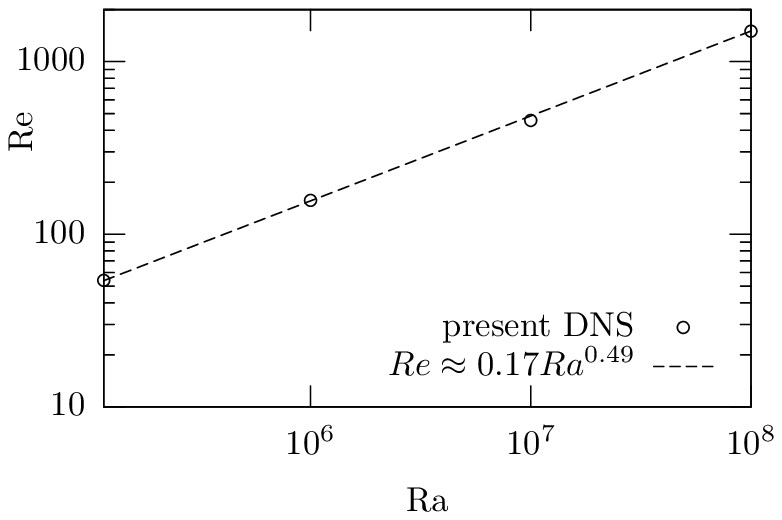}}
  \caption{\label{fig:int_stats}
a) $\Ra$-$\Nu$ scaling for present DNS simulations.
b) $\Re-\Ra$ scaling for present DNS for $\Re$ based on horizontal
squared mean fluctuations, along with a best-fit powerlaw.}
\end{figure*}

Fig. \ref{fig:int_stats}a shows the behavior of $\Nu$ as a function of $\Ra$. 
This result is in good agreement with the relation $\Nu = 0.186\
\Ra^{0.276}$, obtained by DNS with a similar domain and boundary conditions \citep{Kerr1996}, and with the classical wide-aspect ratio experiments of Chu and Goldstein \cite{Chu1973}.
The scaling of $\Re$ as a function of $\Ra$ (Fig. \ref{fig:int_stats}b), where $\Re$ is obtained from the maximum of $\av {\fl{u}\fl{u}}$, has a best-fit scaling as $\Re_u = 0.17\ \Ra^{0.49}$.
This is close to $\Re \propto \Ra^{1/2}$ which corresponds to a Reynolds number
based on the free-fall velocity $U_f = \sqrt{\beta g \Delta \Theta H}$.
Note that the above scaling for $\Re$ is not presumed to describe asymptotic behavior, which cannot be expected in the range of $\Ra$ we consider. Instead it should be treated as a best-fit relation or local exponent.

\section{\label{par:symavresults}Wind-decomposed results}

\subsection{The wind structure}
In order to obtain the realizations for the symmetry-accounting
ensemble-averaging,  the complete three-dimensional fields for $u_i, \Theta$
have been stored twice every convective turnover time, thereby ensuring that the fields are approximately independent.
Furthermore, by performing different simulations at identical $\Ra$ with different initial conditions, a real ensemble averaging was carried out.
The realizations have been selected such that the wind structure has
fully developed \citep{Parodi2004, Roode2004, Hartlep2003}, so that the
criterion for symmetry-accounting ensemble-averaging was satisfied.
Over all ten simulations this resulted in approximately 400 independent
realizations, which were then processed using symmetry-accounted
ensemble-averaging, described in section \ref{par:symav}.

\begin{figure*}
  \center
  \subfigure[]{\includegraphics*[width=80mm]{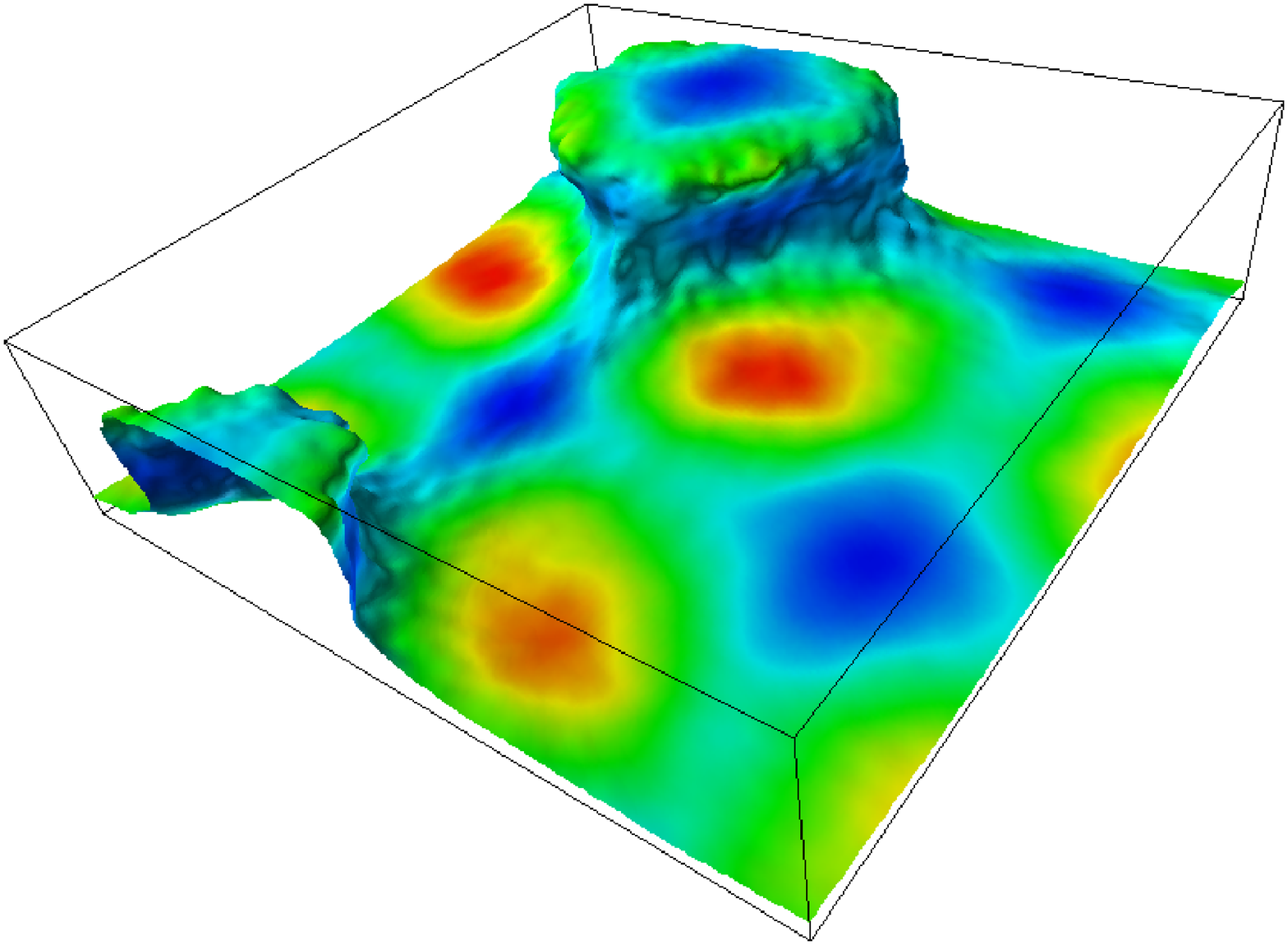}}
  \subfigure[]{\includegraphics*[width=50mm]{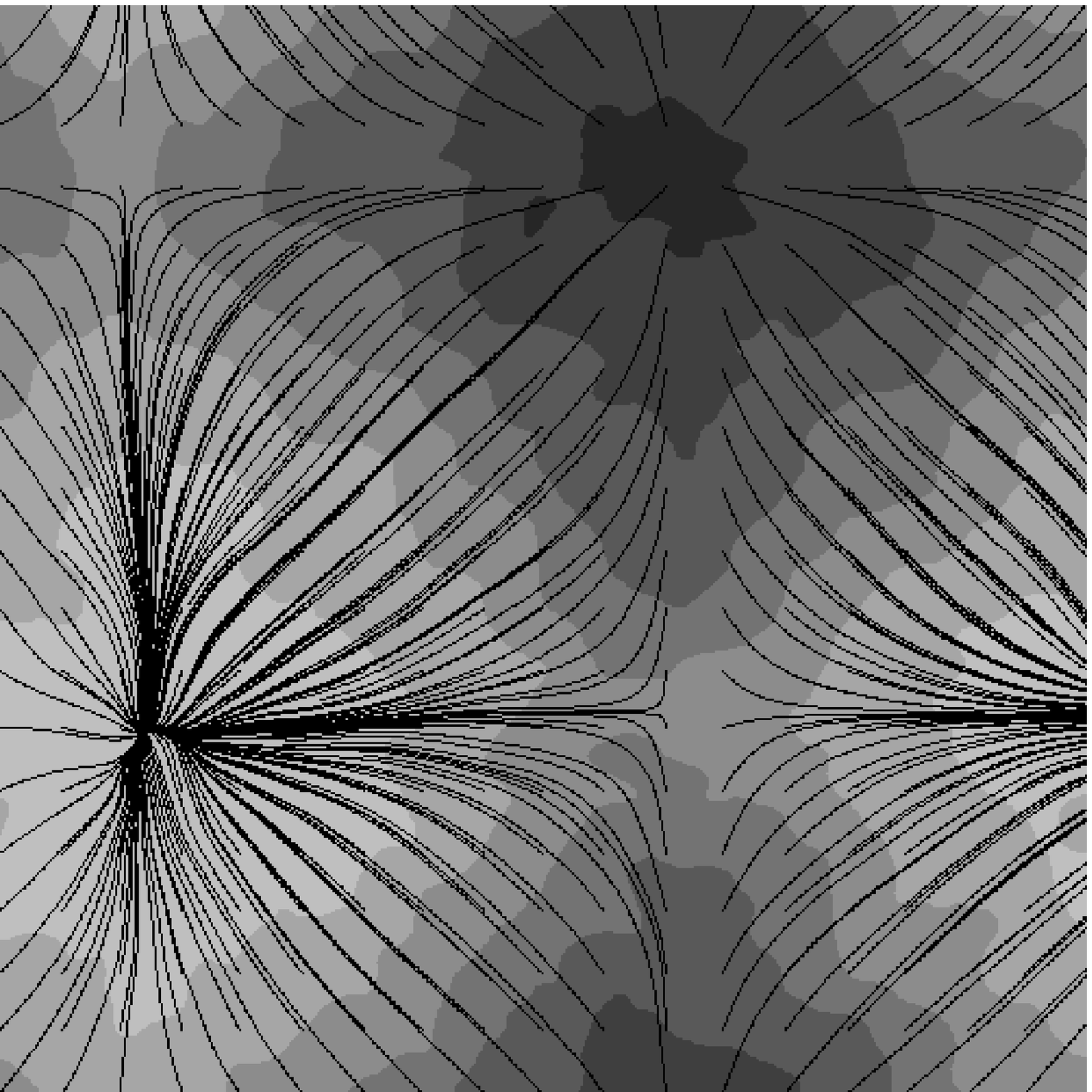}}
  \subfigure[]{\includegraphics*[width=88mm]{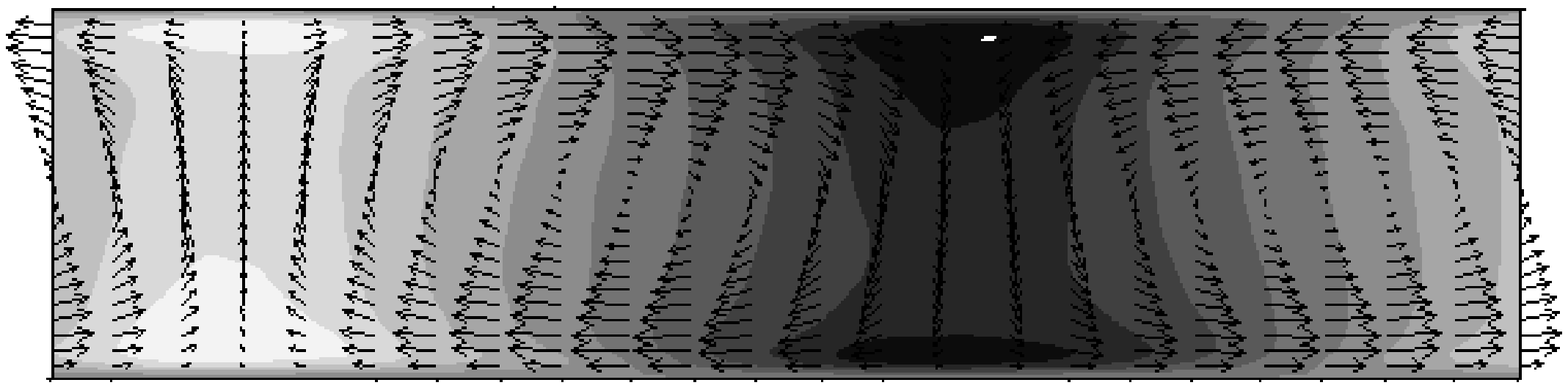}}
  \caption{\label{fig:sav} Results after symmetry-accounted ensemble-averaging
at $\Ra=10^6$ and $\Pr=1$.
a) (color online) 3D iso-surface of temperature, colored by the kinetic energy;
b) Plane-cut in the hydrodynamic boundary layer, iso-contours of relative
temperature $\Theta_r$ and streamlines of the horizontal velocity components;
c) Result after averaging over the $y$-direction (top to bottom in Fig
\ref{fig:sav}b). 
}
\end{figure*}

The result of the averaging is shown in Fig. \ref{fig:sav} for the simulations
at $\Ra=10^6$.
Instead of a one-dimensional temperature profile $\av{\Theta}(z)$, 
a fully three-dimensional temperature field $\sav{\Theta}(x, y, z)$ is obtained 
of which an iso-surface is shown, clearly revealing the wind structure.
These are the fingerprints of the role-like behavior of the wind 
structure.
This is even better visible when making a slice through the hydrodynamic
boundary layer (Fig. \ref{fig:sav}b).
The contour lines are of relative temperature $\sav{\Theta}_r$, which
is the deviation from the plane-averaged temperature $\plav{\sav{\Theta}}(z)$, defined as $\sav{\Theta}_r(x,y,z) \equiv \sav{\Theta}(x,y,z) -
\plav{\sav{\Theta}}(z)$.
The relative temperature $\sav{\Theta}_r$ is closely related to the height-averaged temperature $\hav{\Theta}$ when $\volav{\Theta}=0$, 
as $\hav{\sav{\Theta}_r} = \hav{\Theta}$.
The relative temperature $\sav{\Theta}_r$ is an indicator for where the fluid is rising and falling, as can be seen
from the streamlines of the horizontal components $u,v$.
Figure \ref{fig:sav}c shows a side-view of the average field after averaging
over the y-direction.
Again, the iso-contours are of relative temperature $\sav{\Theta}_r$.
Clearly visible in the figure is the projection of the two rolls onto the side
view.
Note that the periodic boundary conditions rule out the one-roll wind structures that are common for small-aspect ratio cells because of continuity arguments.

\begin{figure}
  \centering
  \setlength{\unitlength}{1mm}
  \includegraphics{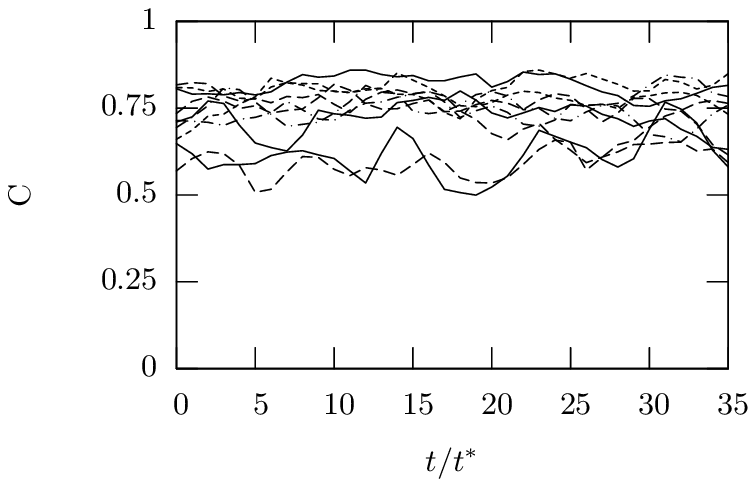}
  \caption{\label{fig:corr}
  Correlation with the wind structure for the 10 simulations at $\Ra=10^6$
and $\Pr=1$ as a function of time. 
  }
\end{figure}

In Fig. \ref{fig:corr}, the correlation of the height-averaged temperature
$\hav{\Theta}$ with the wind structure $\hav{\sav{\Theta}}$ is shown as
a function of time for the ten independent simulations at $\Ra=10^6$.
As will be recalled this is the matching criterion for the symmetry-accounted
average, so the correlation with $\hav{\sav{\Theta}}$ is an indication of how
appropriate the method is, and also for the strength of the wind
structure.
It can be seen that on average, the correlation $C$ with the
wind structure is quite good, fluctuating between $0.5-0.85$ for all
simulations.

\subsection{\label{par:keprofiles}Plane-averaged profiles of kinetic energy}

\begin{figure*}
  \centering
  \subfigure[]{\includegraphics{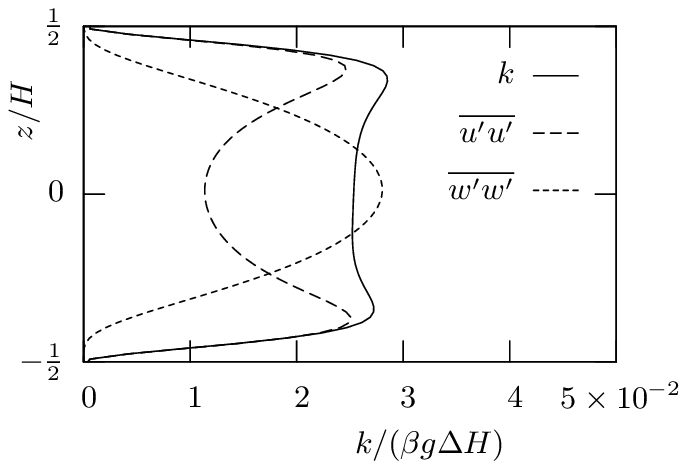}}
  \subfigure[]{\includegraphics{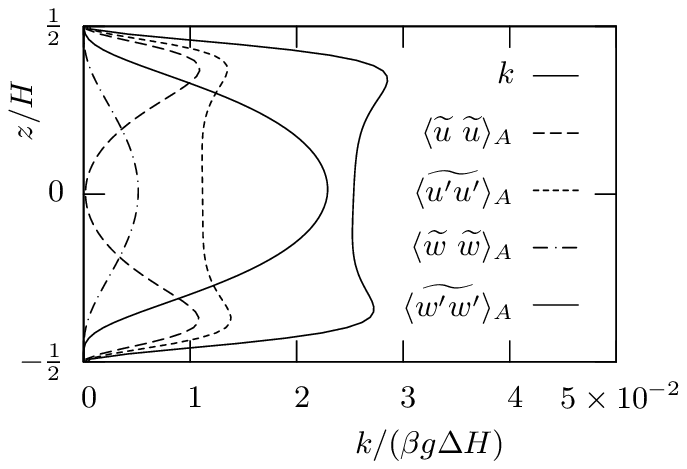}}
  \subfigure[]{\includegraphics{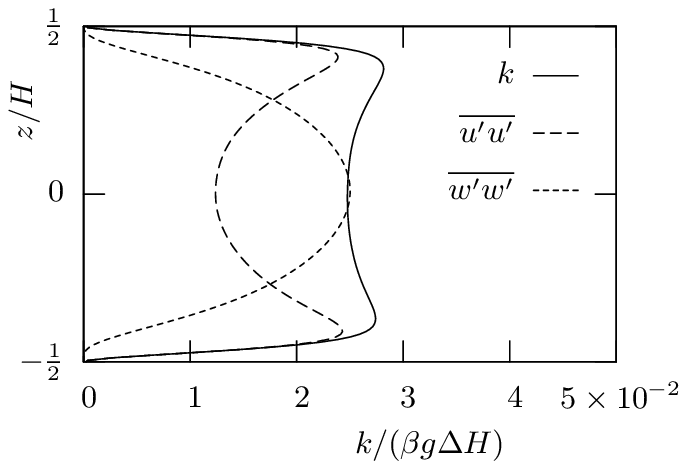}}
  \subfigure[]{\includegraphics{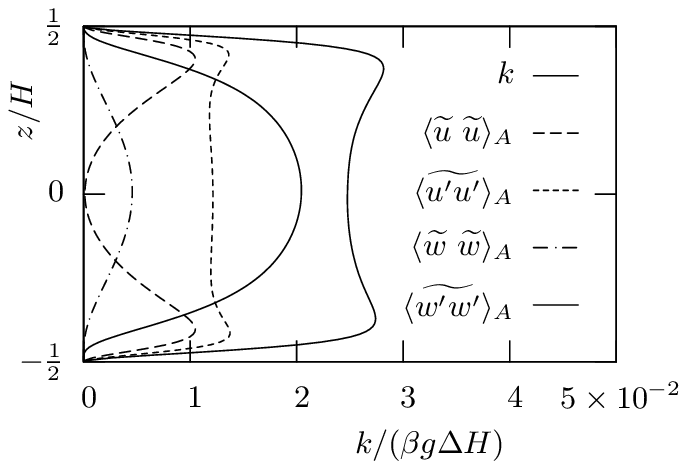}}
  \subfigure[]{\includegraphics{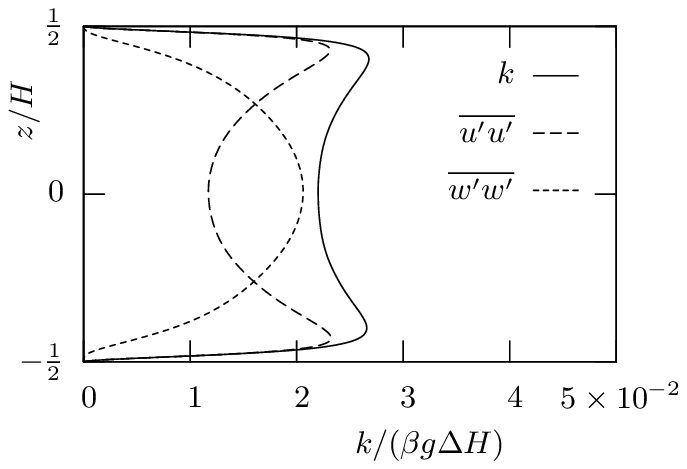}}
  \subfigure[]{\includegraphics{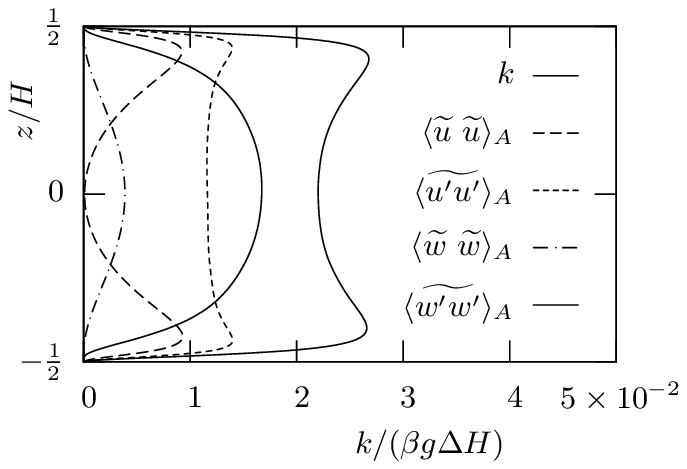}}
  \caption{\label{fig:kprofs}
  Plane-averaged profiles of kinetic energy. Shown are the classical profiles
(a,c,e) and the symmetry-accounted profiles (b,d,f). a,b) $\Ra=1.15 \times 10^5$;
c,d) $\Ra = 1.0 \times 10^6$; e,f) $\Ra=1.0 \times 10^7$.
}
\end{figure*}

Plane-averaged profiles of kinetic energy $k(z)=\plav{\av{\fl{u}_i \fl{u}_i}}$ and its components are shown in Fig. \ref{fig:kprofs}.
Only one of the horizontal components is shown due to homogeneity. 
The classical statistics (Fig.\ \ref{fig:kprofs}a,c,e) only differentiate between the
horizontal and vertical fluctuations as the average velocity $\av{u_i}=0$.
For this reason all variance of the wind structure is transferred to the fluctuations.
From Figs.\ \ref{fig:kprofs}a,c,e one gets an image in which near the bottom 
wall variance of $\av{\fl{u}\fl{u}}$ is created due to the action of the 
plumes impinging on and ejecting from the boundary layers.
The interpretation from the symmetry-accounted profiles (Fig.
\ref{fig:kprofs}b,d,f) is completely different.
Here one sees that the maxima in 
$\av{\fl{u}\fl{u}} = \plav{\sav{u}\sav{u}} + \plav{\sav{\sfl{u}\sfl{u}}}$ are
primarily caused by the wind.
The fluctuations, representing the action of the plumes, are nearly uniformly
distributed in the bulk of the flow, and only a slight increase is visible
near the boundary layers.
The profiles of Fig. \ref{fig:kprofs} scale nearly perfectly with the squared
free-fall velocity $U_f^2 = \beta g \Delta \Theta H$ for all three $\Ra$
numbers.
Note that the plane-averaged momentum flux $\plav{\sav{\sfl{w}\sfl{u}}}$ is not included in Figs.\ \ref{fig:kprofs}a-f, as this term is zero due to the symmetry of the wind structure. 

\subsection{\label{par:heatredistribution}How does the wind affect the
heat transport?}

\begin{figure}
  \centering
  \hspace{6mm}\subfigure[]{\includegraphics{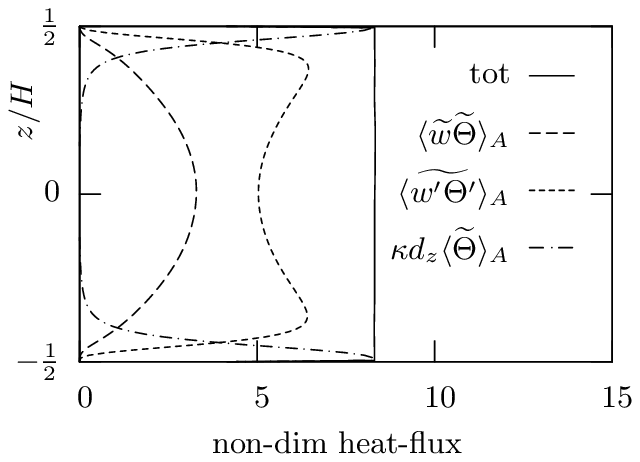}}
  \hspace{-1mm}\subfigure[]{\includegraphics{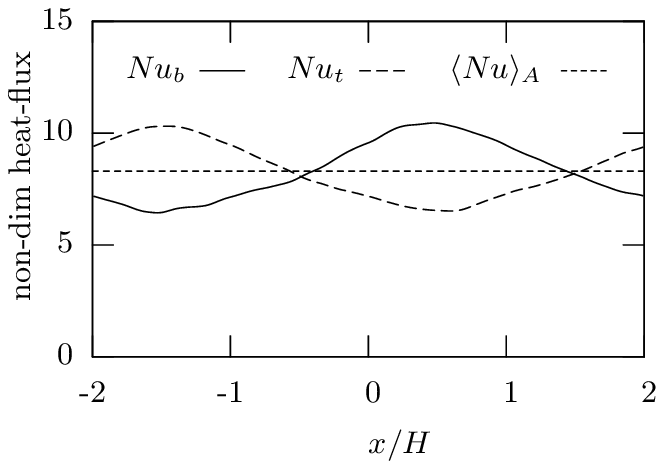}}
  \caption{\label{fig:hf_profs}
  Balance of heat-fluxes at $\Ra=10^6$ and $\Pr=1$.
  a) wind-decomposed heat-fluxes.
  b) $\Nu$ as a function of $x$ and averaged over $y$ at the top and bottom wall.
  }
\end{figure}

Identifying the wind changes the decomposition of the vertical heat-flux. Using \eqref{eq:symav_mean}, \eqref{eq:symav_variance} and the fact that $\plav{w}\equiv 0$ throughout the domain, we can rewrite \eqref{eq:Nu} as:
\begin{equation}
  \Nu = \frac{H}{\kappa \Delta \Theta}
               \left( \plav{\sav{w} \sav{\Theta}} -
                      \plav{\sav{\sfl{w} \sfl{\Theta}}} -
                      \kappa \ddx{z}{\plav{\sav{\Theta}}}
               \right). 
\end{equation}
As $\Nu$ is constant, only the distribution of the three terms on the right-hand side can change as a function of $z$.
This is shown for the simulation at $\Ra=10^6$ in Fig. \ref{fig:hf_profs}a.
Diffusive transport dominates in the boundary layer, where the heat is
transfered to the fluctuations $\sav{\sfl{w}\sfl{\Theta}}$ by
entrainment/detrainment.
In the bulk, about 30 percent of the heat is transported by the wind. 
Here we note that a simple model using sheet plume parameters \citep{Niemela2002} also yields that 30\% of the heat is transported by the mean flow at $\Ra=10^6$.

Where the wind impinges on the wall, the boundary layer will be compressed
and the local $\Nu$ will increase.
Similarly, the local $\Nu$ will decrease in detachment zones.
This effect is demonstrated in Fig. \ref{fig:hf_profs}b where $\Nu$ as a
function of $x$ for the y-averaged wind structure (Fig. \ref{fig:sav}c)
is shown for the top- and bottom wall.
Note that the spatial variations in the wall heat-flux are
generated entirely by the wind structure since $\Nu(x,y) = - \frac{H}{\Delta
\Theta} \ddx{z} {\sav{\Theta}}$ at $z = \pm H/2$.
It can be imagined that spatial variations in $\Nu$ indicate significant
horizontal heat-fluxes as well.
Indeed, this is the case and this point will be addressed below.

\begin{figure}
  \centering
  \hspace{-1cm}
  \subfigure[]{\includegraphics[width=72mm]{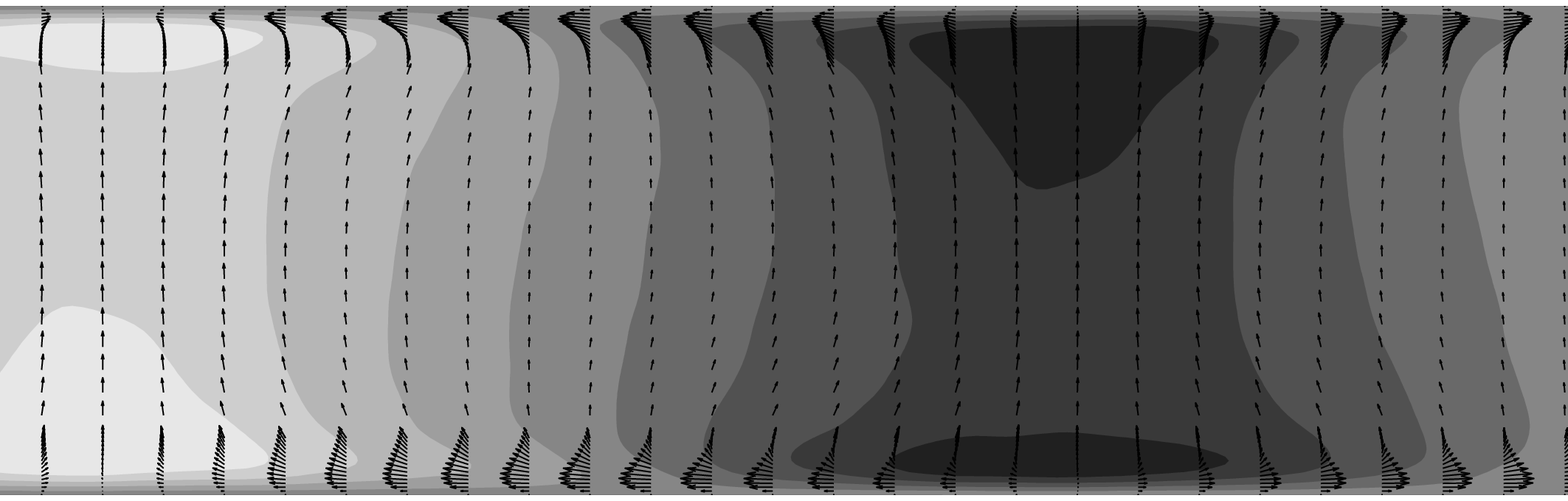}}
  \subfigure[]{\includegraphics{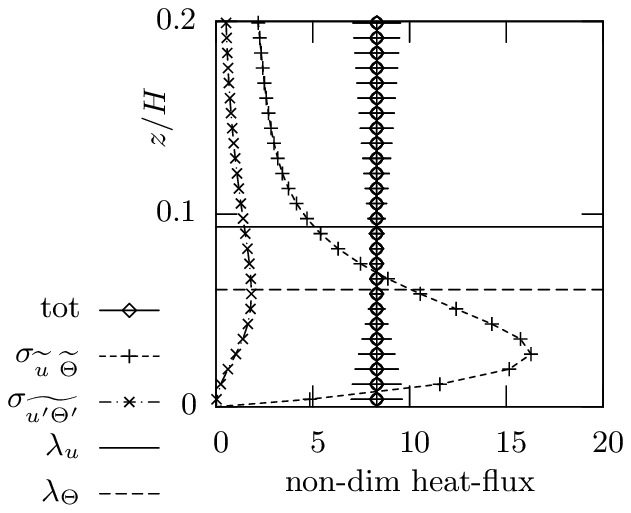}}
  \caption{\label{fig:hf_horz_profs}
  Horizontal heat-fluxes are larger than the vertical and dominate deep in the
thermal boundary layer.
  a) Vectors of the total convective heat-flux $\yav{\sav{u_i} \sav{\Theta}} +
\yav{\sav{\sfl{u_i} \sfl{\Theta}}}$ and iso-contours of relative temperature at
$\Ra=10^6$.
  b) zooming in onto the boundary layer.
  }
\end{figure}

The average horizontal heat-fluxes $\plav{\sav{u_i}\sav{\Theta}}$ and
$\plav{\sav{\sfl{u_i}\sfl{\Theta}}}$ for $i = \{1,2\}$ are zero by definition
due to the absence of a forcing in the horizontal directions.
However, as can be seen in Fig. \ref{fig:hf_horz_profs}a, where the total
convective heat-flux (averaged over the $y$-direction) is shown in flux-vectors $(\sav{u} \sav{\Theta} + \sav{\sfl{u} \sfl{\Theta}}, \sav{w} \sav{\Theta} + \sav{\sfl{w} \sfl{\Theta}})$, the horizontal heat-fluxes are significant, especially very close to the walls.
The heat transport is in the same direction at the top and bottom plates,
and is directed to the relatively hot region where the flow is upward on average.

Due to the anti-symmetry of $\sav{u}\sav{\Theta}$ and $\sav{\sfl{u}\sfl{\Theta}}$ (Fig. \ref{fig:hf_horz_profs}a), their plane-average vanishes.
Hence, $\plav{\sav{u}\sav{\Theta}}$ and $\plav{\sav{\sfl{u}\sfl{\Theta}}}$ cannot be used as an indicator for the strength of the horizontal heat-flux.
However, the spatial standard deviation $\sigma_{\sav{u}\sav{\Theta}}$ and $\sigma_{\sav{\sfl{u}\sfl{\Theta}}}$ are  good indicators, with $\sigma_X$ defined as
\begin{equation}
  \sigma_X = \sqrt{\plav{(X - \plav{X})^2}}.
\end{equation}
The spatial standard deviations (Fig.\ \ref{fig:hf_horz_profs}b) emphasize how close near the wall this heat is transported: The peak of the horizontal heat transfer lies deep inside the thermal boundary layer.
This peak originates purely from the interaction of the mean wind and mean temperature field as $\sav{u} \sav{\Theta}$.
Horizontal heat-fluxes even \emph{exceed} the average vertical heat-fluxes.
These findings emphasize the importance of understanding the boundary layer
structure and its dynamics.

The error bars around the total heat-flux denote the spatial variations in
the total vertical heat-flux $\sav{w}\sav{\Theta}+\sav{\sfl{w}\sfl{\Theta}} -
\kappa \ddx{z}{\sav{\Theta}}$.
An interesting aspect is that these variations are large near the walls (due to
the spatial variations in $\Nu$, see Fig. \ref{fig:hf_profs}b), decrease and
go to a minimum at $z=\lambda_\Theta$, after which the variance
increases again due to the turbulent fluctuations.
This suggests that the thermal boundary acts as a redistributor of heat.

\begin{figure}
  \centering
  \includegraphics{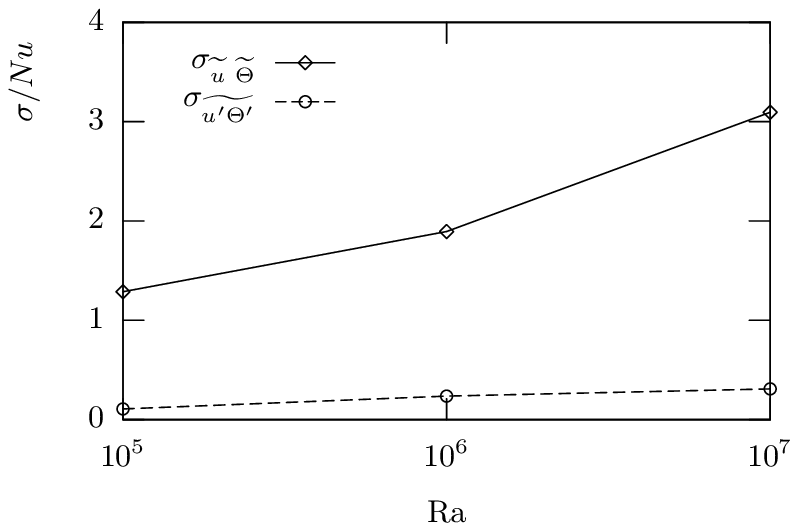}
  \caption{\label{fig:hfx_Ra}
  The peak of spatial standard deviation of the horizontal heat-fluxes
normalized the heat-flux at the wall $\ensav{\phi}|_w$ as a function of $\Ra$,
showing the increasing wind-induced horizontal heat transport.}
\end{figure}

The horizontal heat fluxes become larger as $\Ra$ increases, as shown by the characteristic heat fluxes normalized by $\Nu$ in Fig. \ref{fig:hfx_Ra}.
Shown are the characteristic heat-flux due to the interaction of mean wind and temperature $\sigma_{\sav{u}\sav{\Theta}}$ and turbulent heat flux $\sigma_{\sav{\sfl{u}\sfl{\Theta}}}$.
Although the fluctuations $\sigma_{\sav{\sfl{u}\sfl{\Theta}}}$ grow in strength
relative to $\Nu$ as $\Ra$ increases, their magnitude is still quite small at
$\Ra=10^7$.
In contrast, the heat-flux due to the wind $\sigma_{\sav{u}\sav{\Theta}}$ is
nearly a factor 3 larger than the vertical heat flux at $\Ra=10^7$!
The horizontal heat fluxes are central to the mechanism driving the wind, as is discussed below and in section \ref{par:windmodel}.
\subsection{\label{par:windmechanism}A wind feedback cycle}

In this section we study the momentum- and heat-balances term by term (Table \ref{tab:bdg_mom}).
As the wind structure is statistically in a steady state, the balance is purely a function of space as $\bA+\bD+\bP+\bB+\bR=0$.
Similar to Fig \ref{fig:sav}c, the budget terms have been averaged over the $y$-direction for convenience of presentation.
Several checks were done to ensure that the $y$-averaged momentum budgets are also representative for the three-dimensional field. 

\renewcommand\arraystretch{2}
\renewcommand\tabcolsep{1mm}
\begin{table}
 \begin{center}
\caption{\label{tab:bdg_mom} Budget terms for momentum and heat equation.}
\begin{tabular}{cccccc}
& $\bA$ & $\bD$ & $\bP$ & $\bB$ & $\bR$ \\
$\ddt{\sav{u}_i} =$ & 
    $-\ddxj{\sav{u}_j \sav{u}_i}$ & $+\nu \ddxjsq{\sav{u_i}}$ &
    $-\ddxi{\sav{p}}$ & 
    $+\beta g \sav{\Theta} \delta_{i3}$ & $-\ddxj{\sav{\fl{u}_j \fl{u_i}}}$ \\
$\ddt{\sav{\Theta}} =$ & 
    $-\ddxj{\sav{u}_j \sav{\Theta}}$ & $+\kappa \ddxjsq{\sav{\Theta}}$ &&&
    $-\ddxj{\sav{\fl{u}_j \fl{\Theta}}}$

\end{tabular} 
\end{center}
\end{table}
\renewcommand\arraystretch{1}

\begin{figure*}
  \centering
  \includegraphics[angle=90,width=180mm]{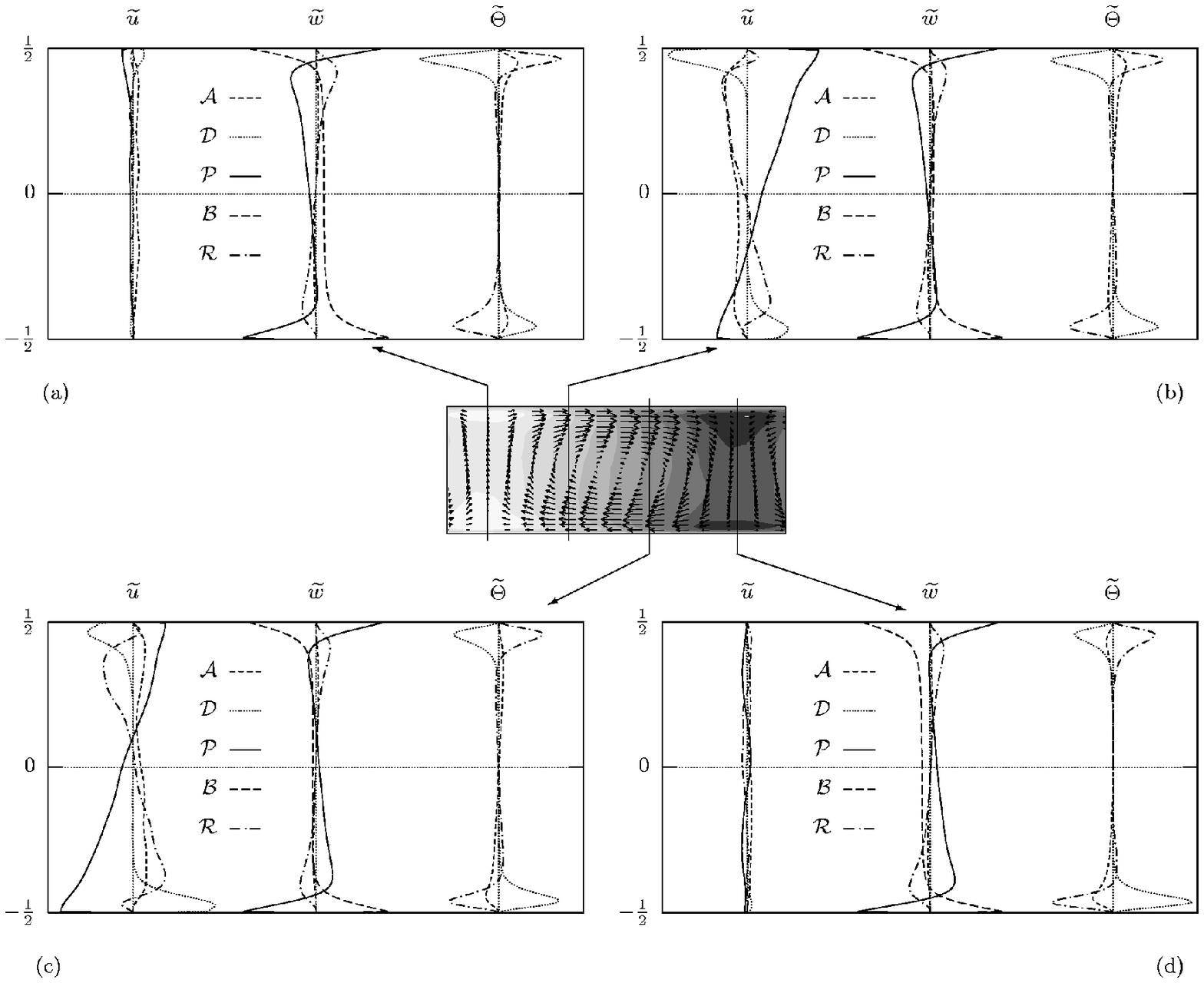}
  \caption{\label{fig:bdg_mom}
  Momentum and temperature budgets as a function of $z$ for $\Ra=10^6$ and $\Pr=1$.
  a) upward motion;
  b) 1/3 of the way;
  c) 2/3 of the way;
  d) downward motion. Note that only half of the wind-structure shown in the center picture (see Fig.\ \ref{fig:sav}). 
  }
\end{figure*}

In Fig. \ref{fig:bdg_mom} four vertical sections are shown, at the location of
maximum upward motion (Fig.\ \ref{fig:bdg_mom}a),
at 1/3 of the cycle (Fig.\ \ref{fig:bdg_mom}b),
at 2/3 of the cycle (Fig.\ \ref{fig:bdg_mom}c) and
at the maximum downward motion (Fig.\ \ref{fig:bdg_mom}d).
Note that this is only half of the flow field; the other half does not provide new information due to symmetry.
In the description it is sufficient to focus on the top wall only, as the top
profiles from Fig.\ \ref{fig:bdg_mom}a can be mapped onto the bottom profiles
from Fig.\ \ref{fig:bdg_mom}d by elementary symmetry operations, and 
the same holds for Fig.\ \ref{fig:bdg_mom}b and Fig.\ \ref{fig:bdg_mom}c.
Focusing on the region where the flow is upward (Fig.\ \ref{fig:bdg_mom}a), the forces of the horizontal momentum equation are nearly zero.
In the vertical momentum equation, the buoyancy term $\bB$ is balanced by the
vertical pressure gradient $\bP$ and the Reynolds stress $\bR$.
In this region, the average temperature is positive, resulting in a positive
buoyancy forcing $\bB$ over nearly the entire vertical.
The vertical pressure gradient is negative with a negative peak near the top
plate which reflects the resulting pressure build-up due to the impinging
plumes.
The Reynolds stresses $\bR$, dominated by the term $-\ddx{z}{\sav{\sfl{w} \sfl{w}}}$, are slightly stronger on the top plate than on the bottom plate.
This is an indication that on average, plume impingement is a more violent 
process than plume detachment.
In the budget for temperature, the balance is primarily between diffusion $\bD$,
gradients in the turbulent heat-flux $\bR$, with a small contribution due to
the local acceleration of the mean flow field $\bA$.
The forcing is stronger at the top plate, due to the impingement of the
wind and the plumes.
The local Nusselt number $\Nu_t$ is maximal at this position (see also Fig.\
\ref{fig:hf_profs}b).
Note that $\Nu_t$ is related to the integral of the thermal diffusive term $\bD$ on the top boundary layer. As the area under $\bD$ at the top-wall is larger than the area under $\bD$ at the bottom wall, it follows that $\Nu_t > \Nu_b$, which is consistent with Fig.\ \ref{fig:hf_profs}.

Following the flow along the top plate, the horizontal momentum budget of Fig.\
\ref{fig:bdg_mom}b shows a strong positive horizontal pressure gradient $\bP$,
which is balanced by diffusion $\bD$ close to the wall, Reynolds stresses
$\bR$ and inertial terms $\bA$ a bit further away.
The horizontal pressure gradient $\bP$ is positive over the upper two thirds of
the vertical.
The interesting small peak in $\bR$ very near the wall will be discussed in more
details in the accompanying paper \cite{vanReeuwijk2007c}, which focuses on the boundary layers.
In the vertical momentum equations, the situation is similar to that of Fig.\
\ref{fig:bdg_mom}a, with the exception that the buoyancy force has become
less positive.
For the temperature budget, $\Nu_t$ is lower at this point here (Fig.\
\ref{fig:hf_profs}b), making thermal diffusion $\bD$ weaker.

A bit further downstream (Fig.\ \ref{fig:bdg_mom}c), the horizontal momentum budgets indicate that the pressure gradient is still positive but has decreased in strength.
As the flow has started to decelerate, the inertial force $\bA$ has a positive
contribution.
Close to the wall, diffusion $\bD$ is braking the fluid, and a bit further away
the fluctuations $\bR$.
As far as the temperature budget is concerned, $\Nu_t$ has decreased even more.
The budgets when the flow comes to a halt and starts its descent down are shown in Fig.\ \ref{fig:bdg_mom}d.
In the vertical momentum equation, the buoyant forcing has become negative
over nearly the entire vertical, which is balanced by the vertical pressure
gradient $\bP$ and the Reynolds stress term $\bR$.
As $\Nu_t$ is at a minimum at this position, thermal diffusion is relatively
small here, and the advective part $\bA$ has become negligible.

Concluding, the mean momentum and temperature budgets show that the wind is driven by pressure gradients.
These pressure gradients are generated as the result of spatial buoyancy differences caused by spatial temperature differences.
This finding is in line with the study by Burr \emph{et al.}\ \citep{Burr2003}, despite the absence of sidewalls.
The pressure gradient can be estimated by integrating the vertical momentum equation, as will be shown in the next section.

\begin{figure}
  \includegraphics[width=85mm]{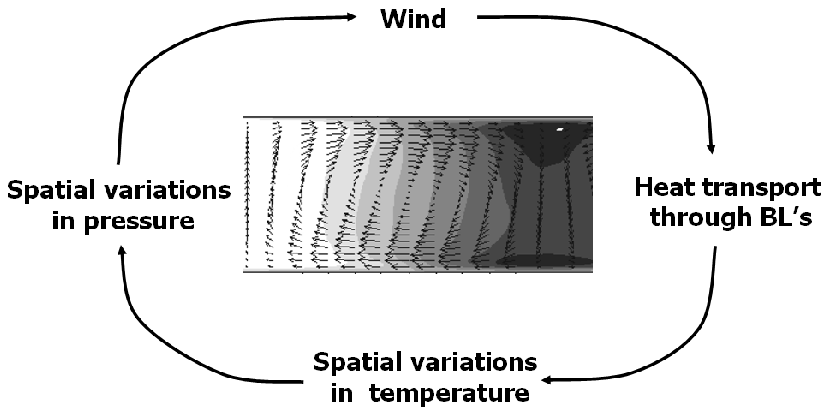}
  \caption{\label{fig:feedback} Wind feedback mechanism}
\end{figure}

Using Fig.\ \ref{fig:bdg_mom} we can identify a detailed feedback mechanism sustaining the wind.
The buoyancy force creates a pressure increase (decrease) on the top wall where the flow is positively (negatively) buoyant.
This generates horizontal pressure gradients at the top- and bottom walls that
drive a mean flow which transports a relatively large amount of heat through the bottom layers (section \ref{par:heatredistribution}).
The net transport of heat towards the region with ascending flow causes spatial temperature gradients (Fig.\ \ref{fig:hf_horz_profs}).
Finally, these spatial temperature differences induce spatial gradients in the
buoyancy which completes the feedback cycle.
A schematic diagram of this process is shown in Fig.\ \ref{fig:feedback}.

\section{\label{par:windmodel}A simple model for the wind}

\subsection{A short derivation}

\begin{figure}
  \centering
  \includegraphics[width=85mm]{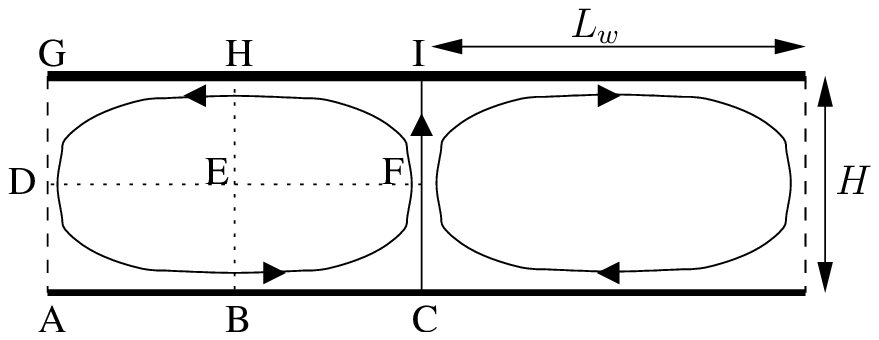}
  \caption{\label{fig:windmodel_CV}Sketch of the wind structure and 9 locations A-I.}
\end{figure}

Based on the feedback mechanism deduced in the previous section, a simple mathematical model can be constructed, by averaging the two-dimensional momentum and temperature equations over appropriate regions of space. A sketch of a typical wind structure is shown in Figure \ref{fig:windmodel_CV}, with 9 locations A-I which will be used to identify specific areas. A generic averaging operator $\spav{\cdot}$, which averages both over lines and areas, is defined as
\begin{gather*}
  \spav{X}_{CI} \equiv \frac{1}{H} \int_{CI} X dz, \\
  \spav{X}_{ACIG} \equiv \frac{1}{H L_w} \iint_{ACIG} X dx dz, 
\end{gather*}
and so on. Here, $L_w$ represents the size of a roll (Fig.\ \ref{fig:windmodel_CV}). As there is a slight clash of variable names (with the height $H$), it should be understood that the locations A-I will only be used as subscripts in the averaging operator.

The model has two main variables, the mean wind velocity $U_w$ and the mean temperature amplitude $\Theta_w$. The mean wind velocity $U_w$ is defined as
\begin{equation}
  \label{eq:Uw}
  U_w \equiv \spav{\sav{u}}_{ACFD} = \frac{2}{H L_w} \iint_{ACFD} \sav{u} dx dz.
\end{equation}
The mean temperature $\Theta_w$ is defined as
\begin{equation}
  \label{eq:Thetaw}
  \Theta_w \equiv \spav{\sav{\Theta}}_{BCIH} 
                = \frac{2}{H L_w} \iint_{BCIH} \sav{\Theta} dx dz,
\end{equation}
which represents the wind-induced temperature amplitude.

Averaging the two-dimensional horizontal momentum equation over the area ACFD and the temperature-equation over the area BCID results in
\begin{align}
 \label{eq:windmodel_Uw_unclosed1}
 \dddt{U_w} &=
 -2 \frac{\spav{\sav{\sfl{w}\sfl{u}}}_{DF}}{H} 
 -\frac{\spav{\sav{p}}_{CF} - \spav{\sav{p}}_{AD}}{L_w}
 - 2 \nu \frac{\spav{\ddx{z}{\sav{u}}}_{AC}}{H}, \\
 \label{eq:windmodel_Thetaw_unclosed1}
  \dddt{\Theta_w} &= 
 \frac{2\spav{\sav{u}\sav{\Theta}}_{BH}}{L_w}
 +\frac{2\spav{\sav{\sfl{u}\sfl{\Theta}}}_{BH}}
       {L_w}
 + \kappa \frac{\spav{\ddx{z}{\sav{\Theta}}}_{HI}
              - \spav{\ddx{z}{\sav{\Theta}}}_{BC}}
               {H}.
\end{align}
A technical discussion about the steps leading to \eqref{eq:windmodel_Uw_unclosed1}, \eqref{eq:windmodel_Thetaw_unclosed1} can be found in appendix \ref{par:windmodel_derivation}. 

\begin{figure}
  \centering
  \subfigure[\label{fig:windmodel_Uw}]{\includegraphics{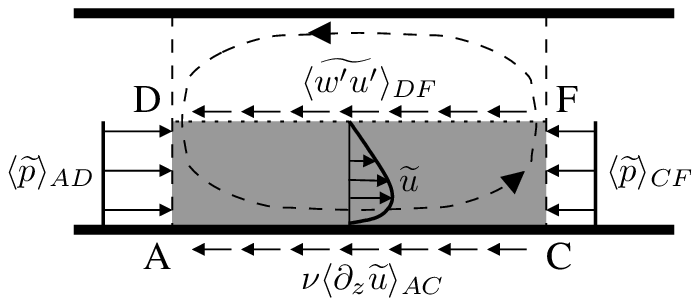}}
  \subfigure[\label{fig:windmodel_Theta}]{\includegraphics{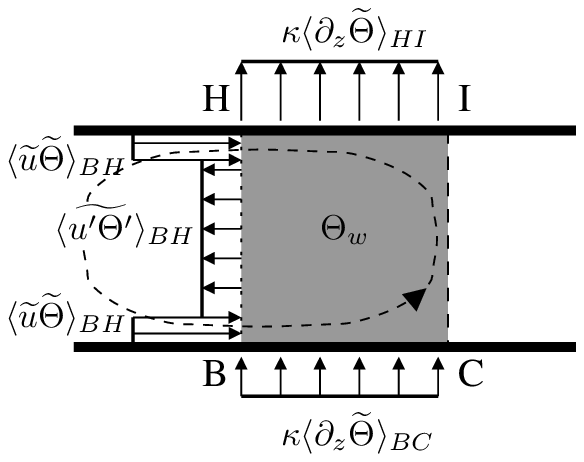}}
  \caption{a) Dominant forces on the wind structure. A pressure gradient drives the wind, while the wall-shear stress and turbulent shear stress in the bulk provide friction; b) Heat fluxes due to the wind structure. The heat-flux $\sav{u}\sav{\Theta}$ creates spatial temperature differences, while the heat-flux at the top and bottom wall and the turbulent heat flux in the bulk destroy temperature differences.}
\end{figure}

In the horizontal momentum equation \eqref{eq:windmodel_Uw_unclosed1}, we see that $U_w$ is driven by a yet unspecified pressure gradient, and is subject to a wall shear stress and a turbulent shear stress in the bulk (see Fig.\ \ref{fig:windmodel_Uw}). 
Both the wall shear stress and the turbulent stress tend to decelerate the wind.
In the heat equation \eqref{eq:windmodel_Thetaw_unclosed1}, the temperature amplitude $\Theta_w$ is driven by  the large horizontal heat flux $\spav{\sav{u} \sav{\Theta}}_{BH}$ in the boundary layer, which was identified in section \ref{par:heatredistribution}. 
The term $\spav{\sav{\sfl{u}\sfl{\Theta}}}_{BH}$ is a horizontal turbulent heat-flux, which tends to decrease temperature differences by turbulent mixing. 
The last term in \eqref{eq:windmodel_Thetaw_unclosed1} represents the heat-flux through the bottom and top wall. If $\Theta_w$ is positive, the heat-flux on the top-wall will be larger than the heat-flux on the bottom wall. Hence, this term effectively removes heat from the control volume. A sketch of the heat-fluxes is shown in Fig.\ \ref{fig:windmodel_Theta}. 

The average pressure gradient can be estimated with the help of the vertical momentum equation. Averaging the vertical momentum equation over CI, which is the streamline connecting the bottom to the top wall, results in
\begin{equation}
\label{eq:windmodel_W1}
\dddt{\spav{w}_{CI}} =  \beta g \spav{\sav{\Theta}}_{CI}
                      - \frac{\spav{\sav{p}}_{I} -\spav{\sav{p}}_{C}}{H}.
\end{equation}
Thus, the average vertical acceleration over CI depends on the average temperature and the pressure difference between the top and the bottom wall. 
Because of the point symmetry around E (Fig.\ \ref{fig:windmodel_CV}), the pressure $\spav{\sav{p}}_I$ is equal to $\spav{\sav{p}}_A$, which means that \eqref{eq:windmodel_W1} provides information about the mean pressure gradient on the bottom wall.
Invoking continuity and approximating the pressure gradient as a linear function of $z$ (see appendix \ref{par:windmodel_derivation}) yields
\begin{equation}
\label{eq:windmodel_presgrad}
\frac{\spav{\sav{p}}_{CF} -\spav{\sav{p}}_{AD}}{L_w} \approx  \frac{H^2}{2L_w^2} \dddt{U_w} - \frac{\beta g H}{2 L_w} \Theta_w.
\end{equation}
This is one of the central results of this paper, as \eqref{eq:windmodel_presgrad} provides an explicit coupling between $U_w$ and $\Theta_w$.

Substituting \eqref{eq:windmodel_presgrad} into \eqref{eq:windmodel_Uw_unclosed1} yields the unclosed equations governing the wind structure:
\begin{align}
 \label{eq:windmodel_Uw_unclosed2}
 \dddt{U_w} &=
\frac{2 L_w^2}{2 L_w^2+H^2} \left(
\frac{\beta g H}{2 L_w} \Theta_w
 -2 \frac{\spav{\sav{\sfl{w}\sfl{u}}}_{DF}}{H} 
 - 2 \nu \frac{\spav{\ddx{z}{\sav{u}}}_{AC}}{H} \right), \\
 \label{eq:windmodel_Thetaw_unclosed2}
  \dddt{\Theta_w} &= 
 \frac{2\spav{\sav{u}\sav{\Theta}}_{BH}}{L_w}
 +\frac{2\spav{\sav{\sfl{u}\sfl{\Theta}}}_{BH}}
       {L_w}
 + \kappa \frac{\spav{\ddx{z}{\sav{\Theta}}}_{HI}
              - \spav{\ddx{z}{\sav{\Theta}}}_{BC}}
               {H}.
\end{align}

\subsection{Parameterization, turbulence closure and dimensionless formulation}

\begin{figure}
  \includegraphics[width=70mm]{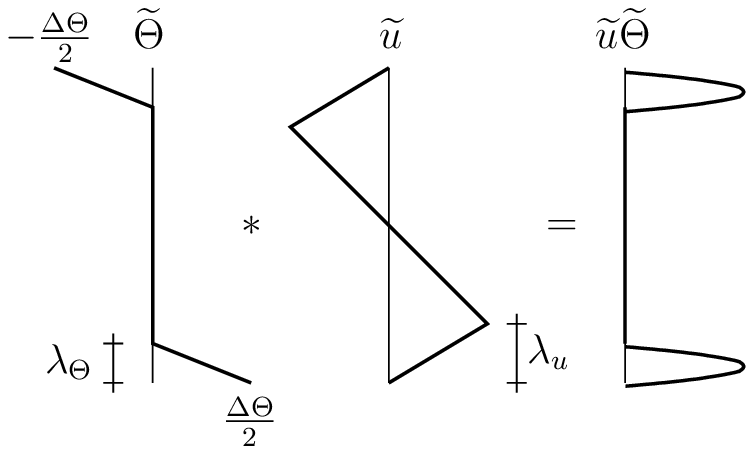}
  \caption{\label{fig:windmodel_utheta}Generation of the horizontal heat-flux $\sav{u}\sav{\Theta}$, which generates spatial temperature differences.}
\end{figure}

The viscous momentum and diffusive heat fluxes at the walls in \eqref{eq:windmodel_Uw_unclosed2},\eqref{eq:windmodel_Thetaw_unclosed2} can be related to $U_w$, $\Theta_w$ and $\lambda_\Theta$ by
\begin{gather*}
 \nu \spav{\ddx{z}{\sav{u}}}_{AC} \approx \frac{1}{2} C_f \abs{U_w} U_w \\
 \kappa \spav{\ddx{z}{\sav{\Theta}}}_{HI} 
     \approx \kappa \frac{-\Delta \Theta / 2 - \Theta_w}{\lambda_\Theta}, \\
 \kappa \spav{\ddx{z}{\sav{\Theta}}}_{BC} 
     \approx \kappa \frac{\Theta_w - \Delta \Theta / 2}{\lambda_\Theta}. \\
\end{gather*}
The wall shear stress $\nu \spav{\ddx{z}{\sav{u}}}_{AC}$ is expressed simply in terms of the friction factor $C_f$ \cite{Schlichting2000}.
The temperature gradient at the top wall $\spav{\ddx{z}{\sav{\Theta}}}_{HI}$ can be estimated by $(-\Delta \Theta / 2 - \Theta_w) / \lambda_\Theta$, as variations in $\lambda_\Theta$ are negligible to first order. 
The temperature gradient at the bottom wall is approximated similarly.
The mean horizontal heat-flux $\spav{\sav{u}\sav{\Theta}}_{BH}$ which drives the flow (section \ref{par:heatredistribution}), is approximated by
\begin{equation*}
 \spav{\sav{u}\sav{\Theta}}_{BH} 
     \approx \frac{\lambda_\Theta U_w \Delta \Theta}{H}.
\end{equation*}
The horizontal heat-flux occurs mainly in the thermal boundary layers (Fig.\ \ref{fig:windmodel_utheta}), where the temperature is approximately $\Delta \Theta / 2$ and the typical velocity is $U_w$. Hence, $\sav{u} \sav{\Theta} \approx U_w \Delta \Theta / 2$, and accounting for the two boundary layer contributions, the average horizontal heatflux $\spav{\sav{u}\sav{\Theta}}_{BH}$ is approximated as above.

The only terms which require closure at this point are the turbulent momentum and heat flux, $\spav{\sav{\sfl{w}\sfl{u}}}_{DF}$ and $\spav{\sav{\sfl{u}\sfl{\Theta}}}_{BH}$ respectively. 
The bulk is well-mixed, as can be judged from the nearly constant temperature and the linearly varying velocity as a function of $z$ in the bulk. 
Therefore, a simple closure with the gradient-diffusion hypothesis is appropriate for the turbulent fluxes
\begin{gather}
  \label{eq:gradhyp_wu}
  \spav{\sav{\sfl{w}\sfl{u}}}_{DF} = - \nu_T \ddx{z}{\sav{u}} 
                                   \approx \nu_T \frac{2 U_w}{H} \\
  \label{eq:gradhyp_uT}
  \spav{\sav{\sfl{u}\sfl{\Theta}}}_{BH} 
     \approx - \kappa_T \spav{\ddx{x}{\sav{\Theta}}}_{BH} 
           = -\frac{\nu_T}{\Pr_T} \frac{2 \Theta_w}{L_w},
\end{gather}
where $\nu_T$ and $\Pr_T=\nu_T/\kappa_T$ are the eddy viscosity and turbulent Prandtl number, respectively. 
To relate $\nu_T$ to mean flow properties, we use the Prandtl mixing length hypothesis, which results in
\begin{equation}
  \label{eq:wind_eddyvisc}
  \nu_T = \alpha \ell^2 \abs{\ddx{z}{u}} \approx \alpha H^2 \frac{\abs{U_w}}{H} = \alpha \abs{U_w} H
\end{equation}
Here $\alpha$ is a free parameter which controls the mixing. 

Using the approximations above, the equations for the wind structure are given by
\begin{gather}
  \label{eq:windmodel_Uw}
  \dddt{U_w} = \frac{2 L_w^2}{2 L_w^2+H^2} \left( 
     \frac{\beta g H}{2 L_w} \Theta_w
    -\frac{4 \alpha + C_f }{H} \abs{U_w} U_w \right) \\
  \label{eq:windmodel_Thetaw}
  \dddt{\Theta_w} = 
 \frac{2 \lambda_\Theta \Delta \Theta}{L_w H} U_w
 -\frac{4 \alpha \abs{U_w} H }
       {L_w^2 \Pr_T} \Theta_w
 - \kappa \frac{2}
               {H \lambda_\Theta} \Theta_w.
\end{gather}

Introducing dimensionless variables $\hat{U}_w = U_w / U_f$, $\hat{\Theta}_w = \Theta_w / \Delta \Theta$, $\hat{t} = t U_f / H$, where $U_f$ is the free-fall velocity $U_f = \sqrt{\beta g H \Delta \Theta}$, results in
\begin{gather}
  \label{eq:windmodel_Uwhat}
  \dddthat{\hat{U}_w} = \frac{2 \hat{L}_w^2}{2 \hat{L}_w^2+1} \left( 
     \frac{1}{2 \hat{L}_w} \hat{\Theta}_w
    -(4 \alpha + C_f) \abs{\hat{U}_w} \hat{U}_w \right), \\
  \label{eq:windmodel_Thetawhat}
  \dddthat{\hat{\Theta}_w} = 
 \frac{2 \hat{\lambda}_\Theta}{\hat{L}_w} \hat{U}_w 
 -\frac{4 \alpha }
       {\hat{L}_w^2 \Pr_T} \abs{\hat{U}_w} \hat{\Theta}_w
 - \frac{2}
               {\hat{\lambda}_\Theta \Re_f \Pr} \hat{\Theta}_w.
\end{gather}
Here, $\hat{L}_w = L_w / H$ and $\hat{\lambda}_{\Theta} = \lambda_\Theta / H$ are the normalised roll size, kinetic and thermal boundary layer thickness. $\Re_f$ is the Reynolds number based upon $U_f$.

The wind model (\ref{eq:windmodel_Uwhat}, \ref{eq:windmodel_Thetawhat}) comprises two nonlinear coupled ordinary differential equations in $\hat{U}_w$ and $\hat{\Theta}_w$. The model contains seven parameters, $\hat{L}_w$, $C_f$, $\hat{\lambda}_{\Theta}$, $\alpha$, $\Re_f$, $\Pr$ and $\Pr_T$. However,  $\hat{\lambda}_{\Theta}=\hat{\lambda}_{\Theta}(\Ra, \Pr)$, $C_f=C_f(\Ra, \Pr)$ and $\Re_f = \Ra^{1/2} \Pr^{-1/2}$. Therefore, the model can be expressed the parameters $\Ra$, $\Pr$, $\hat{L}_w$, $\alpha$ and $\Pr_T$ complemented by the functions for $\lambda_u$ and $C_f$. 
Only $\Pr_T$ and $\alpha$ can be used to calibrate the model, which will be done based on the simulations at $\Ra=10^6$ in the next section.

\subsection{Results}

In this section the model will be compared to the DNS results. As a baseline test, the wind model (\ref{eq:windmodel_Uwhat}, \ref{eq:windmodel_Thetawhat}) should be able to predict the trends in wind speed $\hat{U}_w$ and temperature amplitude $\hat{\Theta}_w$ as a function of $\Ra$. In this study,  we close $\hat{L}_w$, $C_f$ and $\hat{\lambda}_\Theta$ empirically with our DNS results. In particular, we use $\Pr=1$, $\hat{L}_w = 2 \sqrt{2}$, $C_f = A_{\tau} Ra^{\gamma_{\tau}}$ and $\hat{\lambda}_\Theta = A_\Theta \Ra^{\gamma_\Theta}$.
The best-fit coefficients for $C_f$ and $\hat{\lambda}_\Theta$ based on the current simulations are $A_{\tau}=36$, $A_\Theta=2.33$, $\gamma_{\tau}=-0.30$ and $\gamma_\Theta=-0.27$. 

The turbulence parameters $\alpha$ and $\Pr_T$ will be calibrated using the turbulent fluxes and wind and temperature amplitude for the simulation at $\Ra=10^6$.
By calculating $\nu_t$ and $\kappa_T$ with \eqref{eq:gradhyp_wu} and \eqref{eq:gradhyp_uT} it follows that $\Pr_T \approx 0.85$, in reasonable agreement with the generally accepted $\Pr_T \approx 0.9$ for shear flows \cite{Schlichting2000}.
The mixing parameter $\alpha$ can be calculated from \eqref{eq:wind_eddyvisc}, which results in $\alpha \approx 0.6$.
It is noted that $\alpha$ and $\Pr_T$ are not parameters in the strict sense, as the DNS results indicate they have a weak dependence on $\Ra$.

\begin{figure}
  \includegraphics[width=60mm]{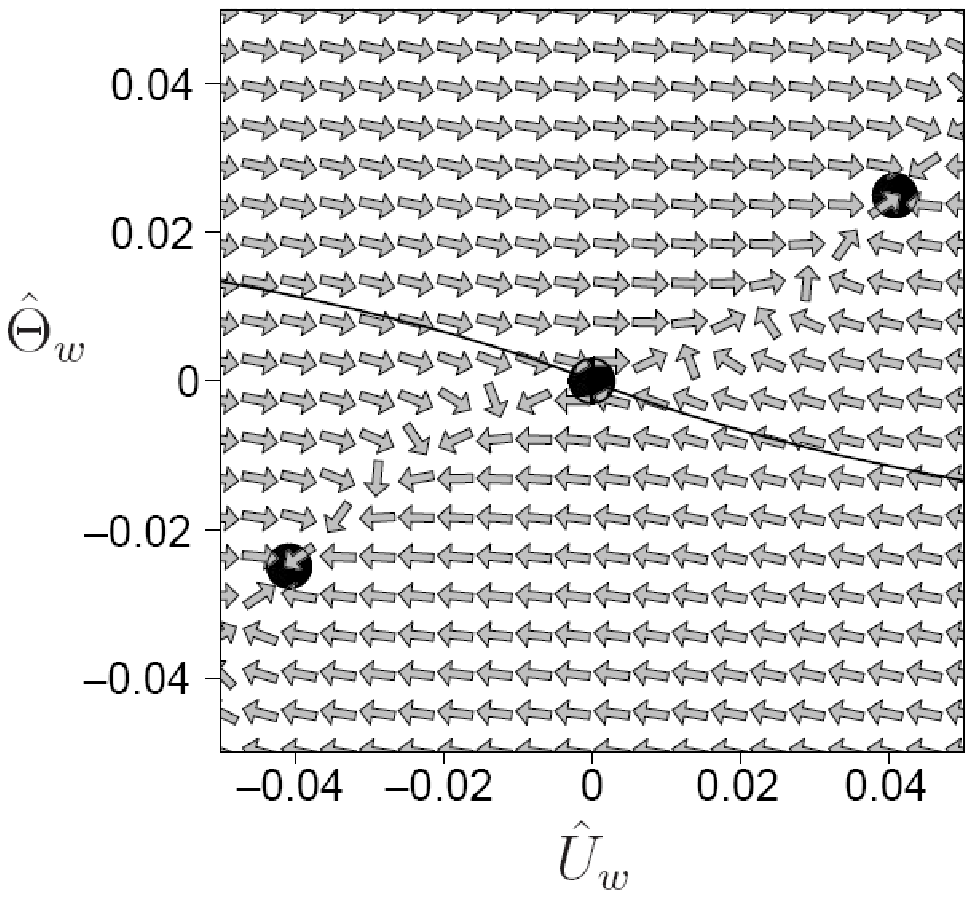}
  \caption{\label{fig:windmodel_phasespace}Phase-space of wind model at $\Ra=10^7$. The fixed points are denoted by circles and the black line is the separatrix.}
\end{figure}

The phase-space of (\ref{eq:windmodel_Uwhat}, \ref{eq:windmodel_Thetawhat}) at $\Ra=10^7$ is shown in Fig.\ \ref{fig:windmodel_phasespace}. There are three fixed points in the domain, of which the one at (0, 0) is a saddle node. The two other fixed points are attractors. 
Thus, if there is no wind initially, any small perturbation caused by turbulent fluctuations will cause the system to settle in a wind structure with either $\hat{U}_w>0$ or $\hat{U}_w<0$.
The tendency of Rayleigh-B\'{e}nard systems to establish a wind structure can thus be explained by the positive feedback created by wind advecting large amounts of heat and the resulting buoyancy differences which drive a mean flow.
The amplitude of the wind is the result of the interaction between the destabilising mechanism mentioned above and the mixing due to turbulence which reduces gradients.
Note that the model cannot describe wind reversals \cite{Sreenivasan2002, Araujo2005, Brown2007}, by the absence of dynamic fluctuations; both nonzero fixed points are stable.
The limitations of the model will be discussed in more details in the concluding remarks (section \ref{par:conclusions}).

As the system is invariant under $\hat{U}_w \rightarrow -\hat{U}_w$, $\hat{\Theta}_w \rightarrow -\hat{\Theta}_w$ it suffices to study the positive fixed point of (\ref{eq:windmodel_Uwhat}, \ref{eq:windmodel_Thetawhat}), which is located at
\begin{gather}
\label{eq:windmodel_Uw_sol}
\hat{U}_w      = \frac{1}{2} \frac{b_3}{b_2} \left(
                 \sqrt{1 + 4 \frac{a_1}{a_2} \frac{b_1 b_2}{b_3^2}}
                - 1  \right) \\
\label{eq:windmodel_Thetaw_sol}
\hat{\Theta}_w = \frac{a_2}{a_1} \hat{U}_w^2
\end{gather}
where
\begin{gather*}
a_1 = \frac{1}{2 \hat{L}_w},~~~
a_2 = 4 \alpha + C_f, \\
b_1 = \frac{2 \hat{\lambda}_\Theta}{\hat{L}_w},~~~ 
b_2 = \frac{4 \alpha }{\hat{L}_w^2 \Pr_T},~~~ 
b_3 = \frac{2}{\hat{\lambda}_\Theta \Re_f \Pr}.
\end{gather*}

\begin{figure}
  \subfigure[\label{fig:windmodel_Uwhat}]
            {\includegraphics[width=75mm]{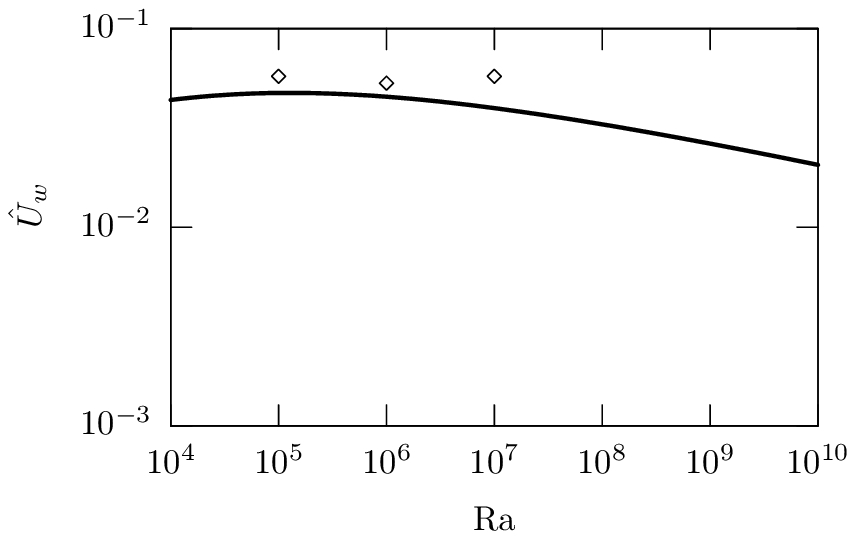}}
  \subfigure[\label{fig:windmodel_Thetawhat}]
            {\includegraphics[width=75mm]{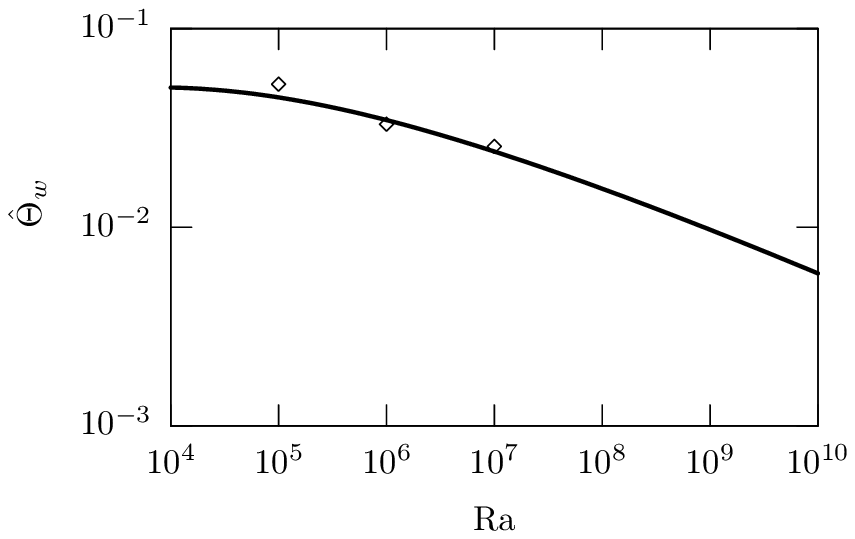}}
  \caption{Behavior of the model (eqns \eqref{eq:windmodel_Uw_sol}, \eqref{eq:windmodel_Thetaw_sol}, solid line) compared to the DNS data (diamonds). a) $\hat{U}_w$ as a function of $\Ra$; b) $\hat{\Theta}_w$ as a function of $\Ra$.}
\end{figure}

Shown in Fig.\ \ref{fig:windmodel_Uwhat} and \ref{fig:windmodel_Thetawhat} are the trends of $\hat{U}_w$ and $\hat{\Theta}_w$ as a function of $\Ra$, compared with the DNS results (diamonds). The model slightly underpredicts $\hat{U}_w$, but the temperature amplitude $\hat{\Theta}_w$ is predicted well. More importantly, the model seems to capture the decreasing trend of $\Theta_w$ properly, as well as very weak $\Ra$ dependence of $\hat{U}_w$. Given its simplicity, the model is in fair agreement with the simulations. 

From \eqref{eq:windmodel_Uw_sol}, it follows that as $\Ra$ increases, $C_f$ becomes negligible relative to the mixing parameter $\alpha$. For the simulation at $\Ra=10^7$, $C_f\approx 0.17$ while $4 \alpha = 2.4$. 
Hence, the friction term $C_f+4\alpha$ is dominated by the turbulence in the bulk.
As $C_f$ is a decreasing function of $\Ra$, this effect becomes stronger as $\Ra$ increases.
This indicates that wall friction has a negligible influence on the wind velocity for $\Ra$ sufficiently high.

The asymptotic scaling of $\hat{U}_w$ for $\Ra \rightarrow \infty$ can be established by studying the scaling of the coefficients of $\hat{U}_w$: 
\begin{align*}
\frac{b_3}{b_2} &= \frac{A_\Theta \hat{L}_w^2 \Pr_T}{2 \alpha A_\Theta} \Ra^{-(1/2 + \gamma_\Theta)} \Pr^{-1/2} \\
4 \frac{a_1}{a_2} \frac{b_1 b_2}{b_3^2} &= \frac{4 \alpha}{4 \alpha+C_f}
                                         \frac {A_\Theta^3}{\hat{L}_w^4~ \Pr_T}
                                         \Ra^{1 + 3 \gamma_\Theta} \Pr
\end{align*}
Assuming that the scaling exponent for $\hat{\lambda}_\Theta$ remains above $\gamma_\Theta = -1/3$, $\Ra^{1+3 \gamma_\Theta} \rightarrow \infty$, by which asymptotic scaling of $\hat{U}_w$ is
\begin{equation}
\hat{U}_w  \propto \Ra^{\gamma_\Theta / 2}
\end{equation}
The wind Reynolds number $\Re_w = U_f H \nu^{-1} U_w / U_f = \hat{U}_w \Ra^{1/2} \Pr^{-1/2}$, so that $\Re_w \propto \Ra^{(1 + \gamma_\Theta)/2}$. 
Based on the exponent from the simulations ($\gamma_\Theta = -0.27$) it follows that $\Re_w \propto \Ra^{0.37}$ in the asymptotic limit. As $\Re_\varepsilon \propto \Ra^{(1- \gamma_\Theta)/3} \propto \Ra^{0.44}$ (where we used that $\lambda_\Theta \propto \Nu^{-1}$), $\Re_w \propto \Ra^{0.37}$ suggests that the wind becomes progressively weaker relative to the fluctuations as $\Ra$ increases.
Naturally one should not assign too much value to the exact exponent, as it critically depends on the $\Ra$ dependence of $\hat{\lambda}_\Theta$. Nevertheless, the flux term generating temperature differences $\sav{u}\sav{\Theta}$ depends critically on $\hat{\lambda}_\Theta$. If $\hat{\lambda}_\Theta$ is a decreasing function of $\Ra$, so will $\Theta_w$ and $U_w$. 
\section{\label{par:conclusions}Concluding remarks}

The aim of this study has been to clarify the processes responsible for the wind amplitude.
Direct numerical simulation was performed at $\Ra=\{ 10^5, 10^6, 10^7,
10^8 \}$ and $\Pr=1$ for an $\Gamma=4$ aspect ratio domain with periodic lateral boundary conditions.
For all but the highest $\Ra$, 10 independent simulations were carried out,
resulting in approximately 400 independent realizations per $\Ra$.
The wind structure was extracted by accounting for symmetries, i.e.\ using the translational invariance of the system to align realizations before averaging them.
In this way, wind could be distinguished from fluctuations for a domain with periodic sidewalls.
It was found that the characteristic peak in the kinetic-energy profile by which the boundary layer thickness is defined, is nearly entirely due to the wind and the turbulent
fluctuations $\plav{\sav{\sfl{u} \sfl{u}}}$ are distributed uniformly outside
the thermal boundary layer.
Deep inside the thermal boundary layers, the wind structure is responsible
for large horizontal heat-fluxes, transporting heat towards the region of upward flow, through the terms $\sav{u}\sav{\Theta}$ and $\sav{v}\sav{\Theta}$.
These horizontal heat-fluxes are up to three times larger than the average
Nusselt number at $\Ra=10^7$, although the total amount of heat transported
through the boundary layer decreases with $\Ra$.
This wind-generated horizontal heat-flux is central for the formation of a wind structure as it generates spatial temperature differences.
As a result of the temperature differences, pressure gradients are generated which drive the wind.

A simple model of two coupled nonlinear ordinary differential equations was derived, which captures the essential processes governing the wind structure.
The primary variables are the wind velocity $U_w$ and the temperature amplitude $\Theta_w$, while the Rayleigh number $\Ra$, the Prandtl number $\Pr$, wind roll size $L_w$, friction factor $C_f(\Ra,\Pr)$ and thermal boundary layer thickness $\lambda_\Theta(\Ra, \Pr)$ are physical parameters.
The turbulence in the bulk is described by a mixing coefficient $\alpha$ and a turbulent Prandtl number $\Pr_T$.
DNS results were used to calibrate $\alpha$ and $\Pr_T$, and served as inspiration for the parameterisation.
The model reproduces the $\Ra$ dependence of $U_w$ and $\Theta_w$ from the DNS, and the following conclusions follow from the wind model:
\begin{itemize}
\item A wind structure is inevitable, as the fixed point corresponding to the absence of wind is an unstable saddle. The positive feedback responsible for this behavior is the interaction between the mean wind and the mean temperature, as described above.
\item The influence of the wall friction on the wind velocity is rather limited. At $\Ra=10^7$, we find that $C_f=0.17$, while $4 \alpha = 2.4$, so that the turbulence in the bulk dominates the total friction $C_f+4 \alpha$ in \eqref{eq:windmodel_Uwhat}.
\end{itemize}

Although the model gives interesting insights, it has a number of limitations.
In the derivation it has been assumed that the domain was unbounded in the lateral directions, i.e.\ no sidewalls.
As a result, the effect of friction on the sidewalls has been omitted, which  -- once included --  will enhance the friction experienced by the wind structure.
Furthermore, the model was derived from the two-dimensional Reynolds-averaged Navier-Stokes equations, which accounts only for the mean effects of the turbulence, thereby excluding long-term dynamical behavior such as reversals and reorientations.
However, no fundamental difficulties are expected to incorporate the missing physics described above.

In the accompanying paper \cite{vanReeuwijk2007c}, we focus on the boundary layers.
Using the wind model developed in this paper, we derive new scaling laws for $\lambda_u$ and $C_f$.
For the wind model, this implies that $\lambda_\Theta$ is the only free parameter in the wind model.
Furthermore, we discuss in detail the issue whether or not the boundary layers should be regarded laminar or turbulent. 

\begin{acknowledgments}
This work is part of the research programme of the Stichting voor Fundamenteel
Onderzoek der Materie (FOM), which is financially supported by the Nederlandse
Organisatie voor Wetenschappelijk Onderzoek (NWO)
The computations were sponsored by the Stichting Nationale Computerfaciliteiten
(NCF).
\end{acknowledgments}

\appendix
\section{\label{par:windmodel_derivation}Derivation of wind model}

In this appendix we average the two-dimensional momentum and temperature equations over specific control volumes in order to develop a theoretical model for the wind. The model has two variables, the wind velocity $U_w$ and the temperature amplitude $\Theta_w$, which are defined in section \ref{par:windmodel}. To identify different regions, various locations are denoted by A-I in Fig.\ \ref{fig:windmodel_CV}. The wind roll size is denoted by $L_w$, and $\spav{\cdot}$ is the generic averaging operator defined in section \ref{par:windmodel}.

\subsection{Horizontal momentum equation}

The two-dimensional horizontal momentum equation is given by
\begin{equation}
\begin{split}
  \ddt{\sav{u}} =& -\ddx{x}{\sav{u}\sav{u}}-\ddx{z}{\sav{w}\sav{u}}
                  -\ddx{x}{\sav{\sfl{u}\sfl{u}}}-\ddx{z}{\sav{\sfl{w}\sfl{u}}}\\
                 & -\ddx{x}{\sav{p}}
                  + \nu (\ddxsq{x}{\sav{u}} + \ddxsq{z}{\sav{u}}).
\end{split}
\end{equation}
This equation will be averaged over the area ACFD, which results in
\begin{equation*}
\begin{split}
  \dddt{U_w} =& 
 -\frac{\overbrace{\spav{\sav{u}\sav{u}}_{CF}}^{=0} 
      - \overbrace{\spav{\sav{u}\sav{u}}_{AD}}^{=0}}{L_w}
 -\frac{\overbrace{\spav{\sav{w}\sav{u}}_{DF}}^{=0} 
      - \overbrace{\spav{\sav{w}\sav{u}}_{AC}}^{=0}}{H/2}
\\ & 
 -\overbrace{\frac{\spav{\sav{\sfl{u}\sfl{u}}}_{CF}
                 - \spav{\sav{\sfl{u}\sfl{u}}}_{AD}}{L_w}}^{\approx 0}
 -\frac{\spav{\sav{\sfl{w}\sfl{u}}}_{DF} 
      - \overbrace{\spav{\sav{\sfl{w}\sfl{u}}}_{AC}}^{=0}}
       {H/2} 
\\ &
 -\frac{\spav{\sav{p}}_{CF} 
      - \spav{\sav{p}}_{AD}}
       {L_w}
\\ &
 + \overbrace{\nu \frac{\spav{\ddx{x}{\sav{u}}}_{CF} 
                      - \spav{\ddx{x}{\sav{u}}}_{AD}}{L_w}}^{\approx 0}
 + \nu \frac{\overbrace{\spav{\ddx{z}{\sav{u}}}_{DF}}^{\approx 0}
           - \spav{\ddx{z}{\sav{u}}}_{AC}}
       {H/2}. 
\end{split}
\end{equation*}
Due to the choice of the control volume, many terms are zero (indicated by $=0$ above them). Other terms can be neglected (indicated by $\approx 0$). The three viscous terms are neglected as they are very small compared to the wall friction term. The average horizontal fluctuations on the interface CF and AD will be approximately of the same strength, so that these terms cancel out. Hence, the horizontal momentum equation simplifies to
\begin{equation}
 \label{eq:windmodel_Uw_unclosed_app}
 \dddt{U_w} =
 -2 \frac{\spav{\sav{\sfl{w}\sfl{u}}}_{DF}}{H} 
 -\frac{\spav{\sav{p}}_{CF} - \spav{\sav{p}}_{AD}}{L_w}
 - 2 \nu \frac{\spav{\ddx{z}{\sav{u}}}_{AC}}{H}.
\end{equation}

\subsection{Temperature equation}

The temperature equation is given by
\begin{equation}
\begin{split}
  \ddt{\sav{\Theta}} =& -\ddx{x}{\sav{u}\sav{\Theta}}-\ddx{z}{\sav{w}\sav{\Theta}}
-\ddx{x}{\sav{\sfl{u}\sfl{\Theta}}}-\ddx{z}{\sav{\sfl{w}\sfl{\Theta}}}\\&
                  + \kappa (\ddxsq{x}{\sav{\Theta}} + \ddxsq{z}{\sav{\Theta}}).
\end{split}
\end{equation}
This equation is averaged over the area BCIH (Fig. \ref{fig:windmodel_CV}), yielding
\begin{equation*}
\begin{split}
  \dddt{\Theta_w} =& 
 -\frac{\overbrace{\spav{\sav{u}\sav{\Theta}}_{CI}}^{=0} 
                 - \spav{\sav{u}\sav{\Theta}}_{BH}}{L_w/2}
 -\frac{\overbrace{\spav{\sav{w}\sav{\Theta}}_{HI}}^{=0} 
      - \overbrace{\spav{\sav{w}\sav{\Theta}}_{BC}}^{=0}}{H}
\\ & 
 -\frac{\overbrace{\spav{\sav{\sfl{u}\sfl{\Theta}}}_{CI}}^{=0}
                 - \spav{\sav{\sfl{u}\sfl{\Theta}}}_{BH}}
       {L_w/2}
 -\frac{\overbrace{\spav{\sav{\sfl{w}\sfl{\Theta}}}_{HI}}^{=0} 
      - \overbrace{\spav{\sav{\sfl{w}\sfl{\Theta}}}_{BC}}^{=0}}
                  {H}
\\ &
 + \overbrace{\kappa \frac{\spav{\ddx{x}{\sav{\Theta}}}_{CI} 
                      - \spav{\ddx{x}{\sav{\Theta}}}_{BH}}{L_w/2}}^{\approx 0}
 + \kappa \frac{\spav{\ddx{z}{\sav{\Theta}}}_{HI}
              - \spav{\ddx{z}{\sav{\Theta}}}_{BC}}
               {H}.
\end{split}
\end{equation*}
Again, the choice of the control volume causes many terms to be zero (indicated by $=0$), while other terms can be neglected (indicated by $ \approx 0$). 
Here, the horizontal diffusive heat-fluxes can be neglected, because they are very small compared to the vertical diffusive heat-fluxes.
The temperature equation is reduced to
\begin{equation}
 \label{eq:windmodel_Thetaw_unclosed_app}
  \dddt{\Theta_w} = 
 \frac{2\spav{\sav{u}\sav{\Theta}}_{BH}}{L_w}
 +\frac{2\spav{\sav{\sfl{u}\sfl{\Theta}}}_{BH}}
       {L_w}
 + \kappa \frac{\spav{\ddx{z}{\sav{\Theta}}}_{HI}
              - \spav{\ddx{z}{\sav{\Theta}}}_{BC}}
               {H}.
\end{equation}

\subsection{Continuity}

The continuity equation
\begin{equation}
  \ddx{x}{\sav{u}} + \ddx{z}{\sav{w}} = 0,
\end{equation}
is averaged over BCFE (Fig.\ \ref{fig:windmodel_CV}), which results in
\begin{equation*}
  \frac{\overbrace{\spav{\sav{u}}_{CF}}^{=0}
       -\spav{\sav{u}}_{BE}}{L_w/2} 
+ \frac{\spav{\sav{w}}_{EF} 
       -\overbrace{\spav{\sav{w}}_{BC}}^{=0}}{H/2}  = 0.
\end{equation*}
Estimating $\spav{\sav{u}}_{BE} \approx U_w$ and $\spav{\sav{w}}_{EF} \approx W_w$, with $W_w$ the mean vertical velocity, the continuity equation becomes
\begin{equation}
  \label{eq:wind_cont}
  \frac{U_w}{L_w} = \frac{W_w}{H}.
\end{equation}

\subsection{Vertical momentum equation}

The unknown pressure gradient can be obtained by averaging the vertical momentum equation over the streamline CI (Fig.\ \ref{fig:windmodel_CV}). As spatial derivatives in the unbounded directions are zero (see Fig.\ \ref{fig:bdg_mom}a), the vertical momentum equation reduces to
\begin{equation}
\ddt{\sav{w}} = \beta g \sav{\Theta}
               - \ddx{z}{\sav{w} \sav{w}}
               + \ddx{z}{\sav{\sfl{w} \sfl{w}}} 
               + \ddx{z}{\sav{p}}
               + \nu \ddxsq{z}{\sav{w}}.
\end{equation}
Averaging over CI gives
\begin{equation*}
\begin{split}
  \dddt{\spav{w}_{CI}} = & \beta g \spav{\sav{\Theta}}_{CI}  
 -\frac{\spav{\sav{p}}_{I} - \spav{\sav{p}}_{C}}{H} \\ &
 -\frac{\overbrace{\spav{\sav{w}\sav{w}}_{I}}^{=0}
      - \overbrace{\spav{\sav{w}\sav{w}}_{C}}^{=0}}{H} 
 -\frac{\overbrace{\spav{\sav{\sfl{w}\sfl{w}}}_{I}}^{=0} 
      - \overbrace{\spav{\sav{\sfl{w}\sfl{w}}}_{C}}^{=0}}{H} \\ & 
 + \nu \frac{\overbrace{\spav{\ddx{z}{\sav{w}}}_{I}}^{=0}
           - \overbrace{\spav{\ddx{z}{\sav{w}}}_{C}}^{=0}}{H}. 
\end{split}
\end{equation*}
It can be verified that $\ddx{z}{\sav{w}} = 0$ at the bottom and top plate by substituting the no-slip boundary condition $u=0$ in the continuity equation. Hence, the average vertical momentum equation reduces to
\begin{equation*}
\dddt{\spav{w}_{CI}} =  \beta g \spav{\sav{\Theta}}_{CI}
                      - \frac{\spav{\sav{p}}_{I} -\spav{\sav{p}}_{C}}{H}.
\end{equation*}
Due to symmetry, the pressure at A and I is identical. Hence, substituting $\spav{\sav{p}}_{I} = \spav{\sav{p}}_{A}$, estimating $\spav{w}_{CI} \approx W_w$, $\spav{\sav{\Theta}}_{CI} \approx \Theta_w$ and using \eqref{eq:wind_cont} gives that the typical pressure gradient at the bottom plate is given by

\begin{equation*}
\frac{\spav{\sav{p}}_{C} -\spav{\sav{p}}_{A}}{L_w} =  \frac{H^2}{L_w^2} \dddt{U_w} - \frac{\beta g H}{L_w} \Theta_w.
\end{equation*}
In Fig. \ref{fig:bdg_mom}, we can see that the pressure gradient is approximately a linear function of $z$, by which the average pressure gradient can be estimated as
\begin{equation}
\label{eq:windmodel_presgrad_app}
\frac{\spav{\sav{p}}_{CF} -\spav{\sav{p}}_{AD}}{L_w} 
    \approx  \frac{H^2}{2 L_w^2} \dddt{U_w} - \frac{\beta g H}{2 L_w} \Theta_w.
\end{equation}
Equations \eqref{eq:windmodel_Uw_unclosed_app}, \eqref{eq:windmodel_Thetaw_unclosed_app} and \eqref{eq:windmodel_presgrad_app} constitute the unclosed wind model.

\bibliography{bib}

\begin{thebibliography}{38}
\expandafter\ifx\csname natexlab\endcsname\relax\def\natexlab#1{#1}\fi
\expandafter\ifx\csname bibnamefont\endcsname\relax
  \def\bibnamefont#1{#1}\fi
\expandafter\ifx\csname bibfnamefont\endcsname\relax
  \def\bibfnamefont#1{#1}\fi
\expandafter\ifx\csname citenamefont\endcsname\relax
  \def\citenamefont#1{#1}\fi
\expandafter\ifx\csname url\endcsname\relax
  \def\url#1{\texttt{#1}}\fi
\expandafter\ifx\csname urlprefix\endcsname\relax\def\urlprefix{URL }\fi
\providecommand{\bibinfo}[2]{#2}
\providecommand{\eprint}[2][]{\url{#2}}

\bibitem[{\citenamefont{Grossmann and Lohse}(2000)}]{Grossmann2000}
\bibinfo{author}{\bibfnamefont{S.}~\bibnamefont{Grossmann}} \bibnamefont{and}
  \bibinfo{author}{\bibfnamefont{D.}~\bibnamefont{Lohse}}, \bibinfo{journal}{J.
  Fluid Mech.} \textbf{\bibinfo{volume}{407}}, \bibinfo{pages}{27}
  (\bibinfo{year}{2000}).

\bibitem[{\citenamefont{Krishnamurti and Howard}(1981)}]{Krishnamurti1981}
\bibinfo{author}{\bibfnamefont{R.}~\bibnamefont{Krishnamurti}}
  \bibnamefont{and} \bibinfo{author}{\bibfnamefont{L.~N.}
  \bibnamefont{Howard}}, \bibinfo{journal}{P. Natl. Acad. Sci. USA}
  \textbf{\bibinfo{volume}{78}}, \bibinfo{pages}{1981} (\bibinfo{year}{1981}).

\bibitem[{\citenamefont{Lam et~al.}(2002)\citenamefont{Lam, Shang, Zhou, and
  Xia}}]{Lam2002}
\bibinfo{author}{\bibfnamefont{S.}~\bibnamefont{Lam}},
  \bibinfo{author}{\bibfnamefont{X.~D.} \bibnamefont{Shang}},
  \bibinfo{author}{\bibfnamefont{S.~Q.} \bibnamefont{Zhou}}, \bibnamefont{and}
  \bibinfo{author}{\bibfnamefont{K.~Q.} \bibnamefont{Xia}},
  \bibinfo{journal}{Phys. Rev. E} \textbf{\bibinfo{volume}{65}},
  \bibinfo{pages}{066306} (\bibinfo{year}{2002}).

\bibitem[{\citenamefont{Niemela et~al.}(2001)\citenamefont{Niemela, Skrbek,
  Sreenivasan, and Donnelly}}]{Niemela2001}
\bibinfo{author}{\bibfnamefont{J.~J.} \bibnamefont{Niemela}},
  \bibinfo{author}{\bibfnamefont{L.}~\bibnamefont{Skrbek}},
  \bibinfo{author}{\bibfnamefont{K.~R.} \bibnamefont{Sreenivasan}},
  \bibnamefont{and} \bibinfo{author}{\bibfnamefont{R.~J.}
  \bibnamefont{Donnelly}}, \bibinfo{journal}{J. Fluid Mech.}
  \textbf{\bibinfo{volume}{449}}, \bibinfo{pages}{169} (\bibinfo{year}{2001}).

\bibitem[{\citenamefont{Qiu and Xia}(1998)}]{Qiu1998}
\bibinfo{author}{\bibfnamefont{X.~L.} \bibnamefont{Qiu}} \bibnamefont{and}
  \bibinfo{author}{\bibfnamefont{K.~Q.} \bibnamefont{Xia}},
  \bibinfo{journal}{Phys. Rev. E} \textbf{\bibinfo{volume}{58}},
  \bibinfo{pages}{486} (\bibinfo{year}{1998}).

\bibitem[{\citenamefont{Qiu et~al.}(2000)\citenamefont{Qiu, Yao, and
  Tong}}]{Qiu2000}
\bibinfo{author}{\bibfnamefont{X.~L.} \bibnamefont{Qiu}},
  \bibinfo{author}{\bibfnamefont{S.~H.} \bibnamefont{Yao}}, \bibnamefont{and}
  \bibinfo{author}{\bibfnamefont{P.}~\bibnamefont{Tong}},
  \bibinfo{journal}{Phys. Rev. E} \textbf{\bibinfo{volume}{61}},
  \bibinfo{pages}{R6075} (\bibinfo{year}{2000}).

\bibitem[{\citenamefont{Sreenivasan et~al.}(2002)\citenamefont{Sreenivasan,
  Bershadskii, and Niemela}}]{Sreenivasan2002}
\bibinfo{author}{\bibfnamefont{K.~R.} \bibnamefont{Sreenivasan}},
  \bibinfo{author}{\bibfnamefont{A.}~\bibnamefont{Bershadskii}},
  \bibnamefont{and} \bibinfo{author}{\bibfnamefont{J.~J.}
  \bibnamefont{Niemela}}, \bibinfo{journal}{Phys. Rev. E}
  \textbf{\bibinfo{volume}{65}}, \bibinfo{pages}{56306} (\bibinfo{year}{2002}).

\bibitem[{\citenamefont{Wang and Xia}(2003)}]{Wang2003}
\bibinfo{author}{\bibfnamefont{J.}~\bibnamefont{Wang}} \bibnamefont{and}
  \bibinfo{author}{\bibfnamefont{K.~Q.} \bibnamefont{Xia}},
  \bibinfo{journal}{Eur. Phys. J. B} \textbf{\bibinfo{volume}{32}},
  \bibinfo{pages}{127} (\bibinfo{year}{2003}).

\bibitem[{\citenamefont{Xi et~al.}(2004)\citenamefont{Xi, Lam, and
  Xia}}]{Xi2004}
\bibinfo{author}{\bibfnamefont{H.~D.} \bibnamefont{Xi}},
  \bibinfo{author}{\bibfnamefont{S.}~\bibnamefont{Lam}}, \bibnamefont{and}
  \bibinfo{author}{\bibfnamefont{K.~Q.} \bibnamefont{Xia}},
  \bibinfo{journal}{J. Fluid Mech.} \textbf{\bibinfo{volume}{503}},
  \bibinfo{pages}{47} (\bibinfo{year}{2004}).

\bibitem[{\citenamefont{Xin and Xia}(1997)}]{Xin1997}
\bibinfo{author}{\bibfnamefont{Y.~B.} \bibnamefont{Xin}} \bibnamefont{and}
  \bibinfo{author}{\bibfnamefont{K.~Q.} \bibnamefont{Xia}},
  \bibinfo{journal}{Phys. Rev. E} \textbf{\bibinfo{volume}{56}},
  \bibinfo{pages}{3010} (\bibinfo{year}{1997}).

\bibitem[{\citenamefont{Xin et~al.}(1996)\citenamefont{Xin, Xia, and
  Tong}}]{Xin1996}
\bibinfo{author}{\bibfnamefont{Y.~B.} \bibnamefont{Xin}},
  \bibinfo{author}{\bibfnamefont{K.~Q.} \bibnamefont{Xia}}, \bibnamefont{and}
  \bibinfo{author}{\bibfnamefont{P.}~\bibnamefont{Tong}},
  \bibinfo{journal}{Phys. Rev. Lett.} \textbf{\bibinfo{volume}{77}},
  \bibinfo{pages}{1266} (\bibinfo{year}{1996}).

\bibitem[{\citenamefont{Kadanoff}(2001)}]{Kadanoff2001}
\bibinfo{author}{\bibfnamefont{L.~P.} \bibnamefont{Kadanoff}},
  \bibinfo{journal}{Phys. Today} \textbf{\bibinfo{volume}{54}},
  \bibinfo{pages}{34} (\bibinfo{year}{2001}),
  \urlprefix\url{http://www.aip.org/pt/vol-54/iss-8/p34.html}.

\bibitem[{\citenamefont{Brown et~al.}(2005)\citenamefont{Brown, Nikolaenko, and
  Ahlers}}]{Brown2005}
\bibinfo{author}{\bibfnamefont{E.}~\bibnamefont{Brown}},
  \bibinfo{author}{\bibfnamefont{A.}~\bibnamefont{Nikolaenko}},
  \bibnamefont{and} \bibinfo{author}{\bibfnamefont{G.}~\bibnamefont{Ahlers}},
  \bibinfo{journal}{Phys. Rev. Lett.} \textbf{\bibinfo{volume}{95}},
  \bibinfo{pages}{084503} (\bibinfo{year}{2005}).

\bibitem[{\citenamefont{Brown and Ahlers}(2006)}]{Brown2006}
\bibinfo{author}{\bibfnamefont{E.}~\bibnamefont{Brown}} \bibnamefont{and}
  \bibinfo{author}{\bibfnamefont{G.}~\bibnamefont{Ahlers}},
  \bibinfo{journal}{J. Fluid Mech.} \textbf{\bibinfo{volume}{568}},
  \bibinfo{pages}{351} (\bibinfo{year}{2006}).

\bibitem[{\citenamefont{Verzicco and Camussi}(2003)}]{Verzicco2003}
\bibinfo{author}{\bibfnamefont{R.}~\bibnamefont{Verzicco}} \bibnamefont{and}
  \bibinfo{author}{\bibfnamefont{R.}~\bibnamefont{Camussi}},
  \bibinfo{journal}{J. Fluid Mech.} \textbf{\bibinfo{volume}{477}},
  \bibinfo{pages}{19} (\bibinfo{year}{2003}).

\bibitem[{\citenamefont{Niemela and Sreenivasan}(2003)}]{Niemela2003}
\bibinfo{author}{\bibfnamefont{J.~J.} \bibnamefont{Niemela}} \bibnamefont{and}
  \bibinfo{author}{\bibfnamefont{K.~R.} \bibnamefont{Sreenivasan}},
  \bibinfo{journal}{J. Fluid Mech.} \textbf{\bibinfo{volume}{481}},
  \bibinfo{pages}{355} (\bibinfo{year}{2003}).

\bibitem[{\citenamefont{Amati et~al.}(2005)\citenamefont{Amati, Koal,
  Massaioli, Sreenivasan, and Verzicco}}]{Amati2005}
\bibinfo{author}{\bibfnamefont{G.}~\bibnamefont{Amati}},
  \bibinfo{author}{\bibfnamefont{K.}~\bibnamefont{Koal}},
  \bibinfo{author}{\bibfnamefont{F.}~\bibnamefont{Massaioli}},
  \bibinfo{author}{\bibfnamefont{K.~R.} \bibnamefont{Sreenivasan}},
  \bibnamefont{and} \bibinfo{author}{\bibfnamefont{R.}~\bibnamefont{Verzicco}},
  \bibinfo{journal}{Phys. Fluids} \textbf{\bibinfo{volume}{17}},
  \bibinfo{pages}{121701} (\bibinfo{year}{2005}).

\bibitem[{\citenamefont{Niemela and Sreenivasan}(2006)}]{Niemela2006}
\bibinfo{author}{\bibfnamefont{J.~J.} \bibnamefont{Niemela}} \bibnamefont{and}
  \bibinfo{author}{\bibfnamefont{K.~R.} \bibnamefont{Sreenivasan}},
  \bibinfo{journal}{J. Fluid Mech.} \textbf{\bibinfo{volume}{557}},
  \bibinfo{pages}{411} (\bibinfo{year}{2006}).

\bibitem[{\citenamefont{Hartlep et~al.}(2005)\citenamefont{Hartlep, Tilgner,
  and Busse}}]{Hartlep2005}
\bibinfo{author}{\bibfnamefont{T.}~\bibnamefont{Hartlep}},
  \bibinfo{author}{\bibfnamefont{A.}~\bibnamefont{Tilgner}}, \bibnamefont{and}
  \bibinfo{author}{\bibfnamefont{F.~H.} \bibnamefont{Busse}},
  \bibinfo{journal}{J. Fluid Mech.} \textbf{\bibinfo{volume}{544}},
  \bibinfo{pages}{309} (\bibinfo{year}{2005}).

\bibitem[{\citenamefont{Shishkina and Wagner}(2006)}]{Shishkina2006}
\bibinfo{author}{\bibfnamefont{O.}~\bibnamefont{Shishkina}} \bibnamefont{and}
  \bibinfo{author}{\bibfnamefont{C.}~\bibnamefont{Wagner}},
  \bibinfo{journal}{J. Fluid Mech.} \textbf{\bibinfo{volume}{546}},
  \bibinfo{pages}{51} (\bibinfo{year}{2006}).

\bibitem[{\citenamefont{Kerr}(1996)}]{Kerr1996}
\bibinfo{author}{\bibfnamefont{R.~M.} \bibnamefont{Kerr}}, \bibinfo{journal}{J.
  Fluid Mech.} \textbf{\bibinfo{volume}{310}}, \bibinfo{pages}{139}
  (\bibinfo{year}{1996}).

\bibitem[{\citenamefont{Verdoold et~al.}(2006)\citenamefont{Verdoold, Tummers,
  and Hanjali\'{c}}}]{Verdoold2006}
\bibinfo{author}{\bibfnamefont{J.}~\bibnamefont{Verdoold}},
  \bibinfo{author}{\bibfnamefont{M.~J.} \bibnamefont{Tummers}},
  \bibnamefont{and}
  \bibinfo{author}{\bibfnamefont{K.}~\bibnamefont{Hanjali\'{c}}},
  \bibinfo{journal}{Phys. Rev. E} \textbf{\bibinfo{volume}{73}},
  \bibinfo{pages}{056304} (\bibinfo{year}{2006}).

\bibitem[{\citenamefont{{Fontenele Araujo}
  et~al.}(2005)\citenamefont{{Fontenele Araujo}, Grossmann, and
  Lohse}}]{Araujo2005}
\bibinfo{author}{\bibfnamefont{F.}~\bibnamefont{{Fontenele Araujo}}},
  \bibinfo{author}{\bibfnamefont{S.}~\bibnamefont{Grossmann}},
  \bibnamefont{and} \bibinfo{author}{\bibfnamefont{D.}~\bibnamefont{Lohse}},
  \bibinfo{journal}{Phys. Rev. Lett.} \textbf{\bibinfo{volume}{9508}},
  \bibinfo{pages}{4502} (\bibinfo{year}{2005}).

\bibitem[{\citenamefont{Brown and Ahlers}(2007)}]{Brown2007}
\bibinfo{author}{\bibfnamefont{E.}~\bibnamefont{Brown}} \bibnamefont{and}
  \bibinfo{author}{\bibfnamefont{G.}~\bibnamefont{Ahlers}},
  \bibinfo{journal}{Phys. Rev. Lett.} \textbf{\bibinfo{volume}{98}},
  \bibinfo{pages}{134501} (\bibinfo{year}{2007}).

\bibitem[{\citenamefont{Burr et~al.}(2003)\citenamefont{Burr, Kinzelbach, and
  Tsinober}}]{Burr2003}
\bibinfo{author}{\bibfnamefont{U.}~\bibnamefont{Burr}},
  \bibinfo{author}{\bibfnamefont{W.}~\bibnamefont{Kinzelbach}},
  \bibnamefont{and} \bibinfo{author}{\bibfnamefont{A.}~\bibnamefont{Tsinober}},
  \bibinfo{journal}{Phys. Fluids} \textbf{\bibinfo{volume}{15}},
  \bibinfo{pages}{2313} (\bibinfo{year}{2003}).

\bibitem[{\citenamefont{van Reeuwijk et~al.}(2007)\citenamefont{van Reeuwijk,
  Jonker, and Hanjali\'{c}}}]{vanReeuwijk2007c}
\bibinfo{author}{\bibfnamefont{M.}~\bibnamefont{van Reeuwijk}},
  \bibinfo{author}{\bibfnamefont{H.~J.~J.} \bibnamefont{Jonker}},
  \bibnamefont{and}
  \bibinfo{author}{\bibfnamefont{K.}~\bibnamefont{Hanjali\'{c}}},
  \bibinfo{journal}{Submitted to Phys. Rev. E}  (\bibinfo{year}{2007}),
  \urlprefix\url{http://arxiv.org/abs/0709.1891}.

\bibitem[{\citenamefont{Hartlep et~al.}(2003)\citenamefont{Hartlep, Tilgner,
  and Busse}}]{Hartlep2003}
\bibinfo{author}{\bibfnamefont{T.}~\bibnamefont{Hartlep}},
  \bibinfo{author}{\bibfnamefont{A.}~\bibnamefont{Tilgner}}, \bibnamefont{and}
  \bibinfo{author}{\bibfnamefont{F.~H.} \bibnamefont{Busse}},
  \bibinfo{journal}{Phys. Rev. Lett.} \textbf{\bibinfo{volume}{91}},
  \bibinfo{pages}{064501} (\bibinfo{year}{2003}).

\bibitem[{\citenamefont{de~Roode et~al.}(2004)\citenamefont{de~Roode,
  Duynkerke, and Jonker}}]{Roode2004}
\bibinfo{author}{\bibfnamefont{S.~R.} \bibnamefont{de~Roode}},
  \bibinfo{author}{\bibfnamefont{P.~G.} \bibnamefont{Duynkerke}},
  \bibnamefont{and} \bibinfo{author}{\bibfnamefont{H.~J.~J.}
  \bibnamefont{Jonker}}, \bibinfo{journal}{J. Atmos. Sci.}
  \textbf{\bibinfo{volume}{61}}, \bibinfo{pages}{403} (\bibinfo{year}{2004}).

\bibitem[{\citenamefont{van Reeuwijk et~al.}(2005)\citenamefont{van Reeuwijk,
  Jonker, and Hanjali\'{c}}}]{vanReeuwijk2005}
\bibinfo{author}{\bibfnamefont{M.}~\bibnamefont{van Reeuwijk}},
  \bibinfo{author}{\bibfnamefont{H.~J.~J.} \bibnamefont{Jonker}},
  \bibnamefont{and}
  \bibinfo{author}{\bibfnamefont{K.}~\bibnamefont{Hanjali\'{c}}},
  \bibinfo{journal}{Phys. Fluids} \textbf{\bibinfo{volume}{17}},
  \bibinfo{pages}{051704} (\bibinfo{year}{2005}).

\bibitem[{\citenamefont{Frisch}(1995)}]{Frisch1995}
\bibinfo{author}{\bibfnamefont{U.}~\bibnamefont{Frisch}},
  \emph{\bibinfo{title}{Turbulence}} (\bibinfo{publisher}{Cambridge University
  Press}, \bibinfo{year}{1995}).

\bibitem[{\citenamefont{Galanti and Tsinober}(2004)}]{Galanti2004}
\bibinfo{author}{\bibfnamefont{B.}~\bibnamefont{Galanti}} \bibnamefont{and}
  \bibinfo{author}{\bibfnamefont{A.}~\bibnamefont{Tsinober}},
  \bibinfo{journal}{Phys. Lett. A} \textbf{\bibinfo{volume}{330}},
  \bibinfo{pages}{173} (\bibinfo{year}{2004}).

\bibitem[{\citenamefont{Eiff and Keffer}(1997)}]{Eiff1997}
\bibinfo{author}{\bibfnamefont{O.~S.} \bibnamefont{Eiff}} \bibnamefont{and}
  \bibinfo{author}{\bibfnamefont{J.~F.} \bibnamefont{Keffer}},
  \bibinfo{journal}{J. Fluid Mech.} \textbf{\bibinfo{volume}{333}},
  \bibinfo{pages}{161} (\bibinfo{year}{1997}).

\bibitem[{\citenamefont{Verstappen and Veldman}(2003)}]{Verstappen2003}
\bibinfo{author}{\bibfnamefont{R.~W. C.~P.} \bibnamefont{Verstappen}}
  \bibnamefont{and} \bibinfo{author}{\bibfnamefont{A.~E.~P.}
  \bibnamefont{Veldman}}, \bibinfo{journal}{J. Comput. Phys.}
  \textbf{\bibinfo{volume}{187}}, \bibinfo{pages}{343} (\bibinfo{year}{2003}).

\bibitem[{\citenamefont{van Reeuwijk}(2007)}]{vanReeuwijk2007}
\bibinfo{author}{\bibfnamefont{M.}~\bibnamefont{van Reeuwijk}}, Ph.D. thesis,
  \bibinfo{school}{Delft University of Technology} (\bibinfo{year}{2007}),
  \urlprefix\url{http://repository.tudelft.nl/file/525273/372306}.

\bibitem[{\citenamefont{Chu and Goldstein}(1973)}]{Chu1973}
\bibinfo{author}{\bibfnamefont{T.~Y.} \bibnamefont{Chu}} \bibnamefont{and}
  \bibinfo{author}{\bibfnamefont{R.~J.} \bibnamefont{Goldstein}},
  \bibinfo{journal}{J. Fluid Mech.} \textbf{\bibinfo{volume}{60}},
  \bibinfo{pages}{141} (\bibinfo{year}{1973}).

\bibitem[{\citenamefont{Parodi et~al.}(2004)\citenamefont{Parodi, von
  Hardenberg, Passoni, Provenzale, and Spiegel}}]{Parodi2004}
\bibinfo{author}{\bibfnamefont{A.}~\bibnamefont{Parodi}},
  \bibinfo{author}{\bibfnamefont{J.}~\bibnamefont{von Hardenberg}},
  \bibinfo{author}{\bibfnamefont{G.}~\bibnamefont{Passoni}},
  \bibinfo{author}{\bibfnamefont{A.}~\bibnamefont{Provenzale}},
  \bibnamefont{and} \bibinfo{author}{\bibfnamefont{E.~A.}
  \bibnamefont{Spiegel}}, \bibinfo{journal}{Phys. Rev. Lett.}
  \textbf{\bibinfo{volume}{92}}, \bibinfo{pages}{194503}
  (\bibinfo{year}{2004}).

\bibitem[{\citenamefont{Niemela and Sreenivasan}(2002)}]{Niemela2002}
\bibinfo{author}{\bibfnamefont{J.~J.} \bibnamefont{Niemela}} \bibnamefont{and}
  \bibinfo{author}{\bibfnamefont{K.~R.} \bibnamefont{Sreenivasan}},
  \bibinfo{journal}{Physica A} \textbf{\bibinfo{volume}{315}},
  \bibinfo{pages}{203} (\bibinfo{year}{2002}).

\bibitem[{\citenamefont{Schlichting and Gersten}(2000)}]{Schlichting2000}
\bibinfo{author}{\bibfnamefont{H.}~\bibnamefont{Schlichting}} \bibnamefont{and}
  \bibinfo{author}{\bibfnamefont{K.}~\bibnamefont{Gersten}},
  \emph{\bibinfo{title}{Boundary layer theory}}
  (\bibinfo{publisher}{McGraw-Hill}, \bibinfo{year}{2000}).

\end{thebibliography}
\bibliographystyle{apsrev}

\end{document}